\documentclass[a4paper,12pt,twoside]{report}
\usepackage[left=2cm,right=2cm,top=2cm,bottom=3cm]{geometry}
\pdfoutput=1
\usepackage{amsmath}
\usepackage{bigints}
\usepackage{amssymb}
\usepackage{relsize}
\usepackage{graphicx}
\usepackage{array}
\usepackage{tabu}
\usepackage{caption}
\graphicspath{{Figures/}}
\usepackage{subcaption}
\usepackage{float}
\usepackage{hyperref}
\usepackage{xcolor}
\hypersetup{
colorlinks = true,
urlcolor = red,
linkcolor = blue,
citecolor = blue
}

\begin{document}

\begin{titlepage}

\begin{center}
\vspace*{2cm}

\bf{\LARGE Dark Matter Self Interactions and its Impact on Large Scale Structures}\\
\vspace*{2cm}

{\Large
{\bf Chayan Chatterjee}\\
\vspace*{4mm}
\begin{small}
{\sl under the supervision of}
\end{small} \\
\vspace*{10mm}
{\bf Prof. Debasish Majumdar\\
Saha Institute of Nuclear Physics}\\
\begin{small}
{\sl and}
\end{small}\\
\vspace*{3mm}
{\bf Dr. Suchetana Chatterjee\\
Presidency University}\\
}

\vspace{2cm}
\begin{figure}[h]
\centering
\includegraphics[width=0.3\textwidth]{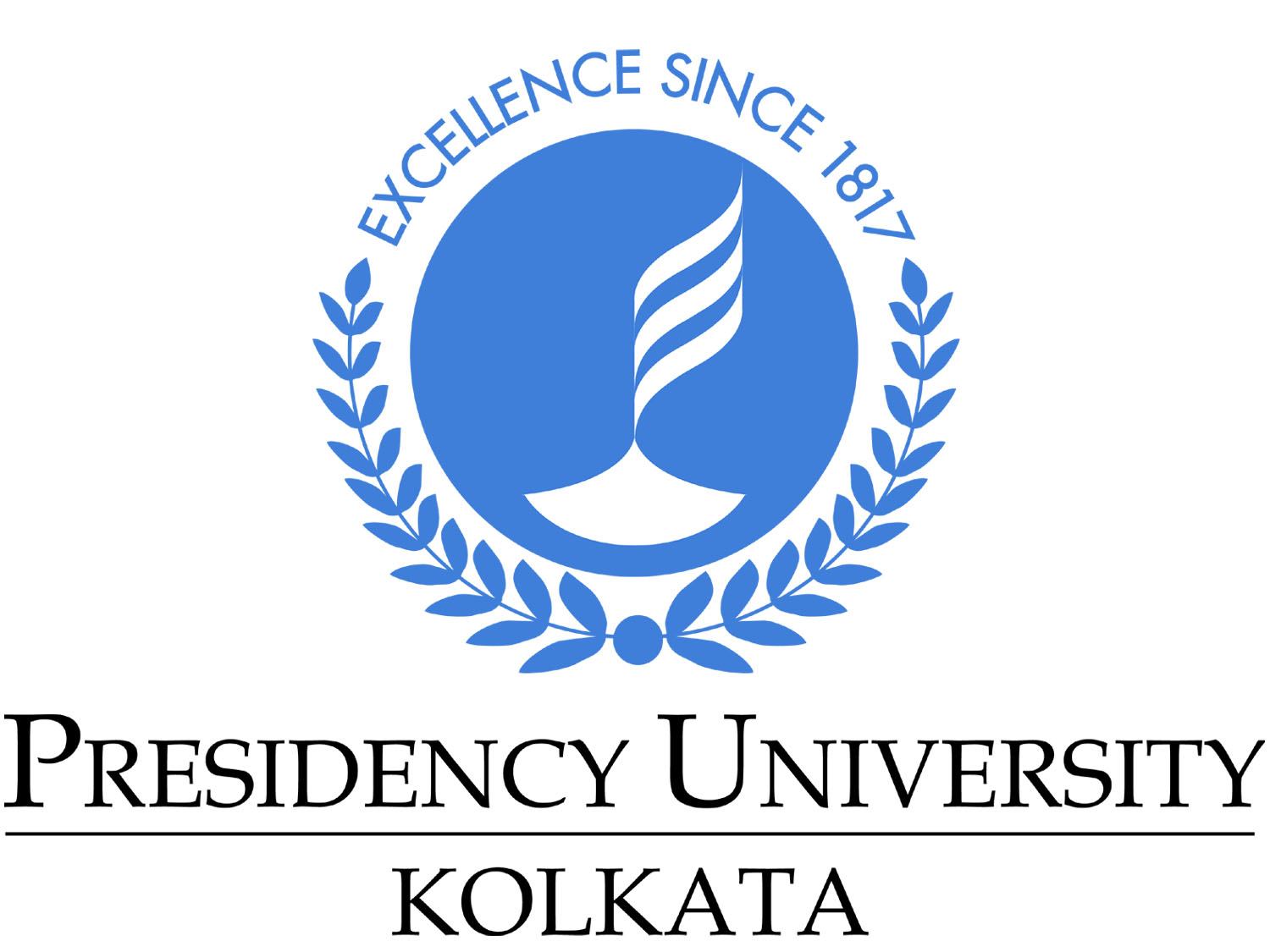}
\end{figure}
\vspace{1.5cm}

A thesis submitted in partial fulfilment of the requirements for the degree of Master of Science in Physics at Presidency University, Kolkata.\\
\vspace{1cm}
May 2018

\end{center}

\end{titlepage}

%\preface
\newpage
\thispagestyle{empty}
%\vspace*{0.1cm}
%\vskip
\begin{center}
\resizebox{5cm}{!}{\includegraphics[angle=0]{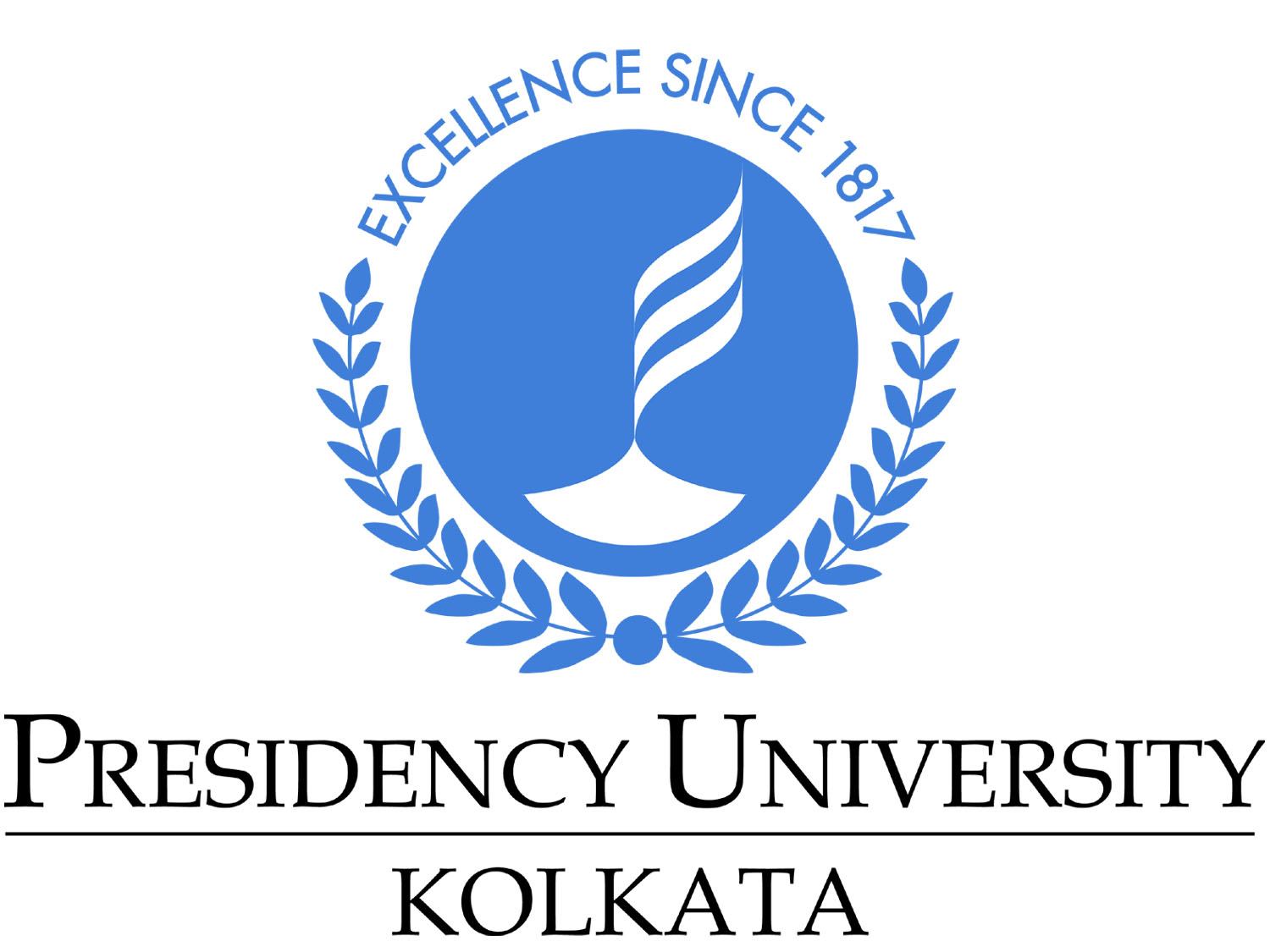}} \\
\vspace{1.7cm}
\begin{Large}
\textbf{CERTIFICATE}\\
\end{Large}
\end{center}
\vspace{0.7cm}
\begin{large}
This is to certify that Chayan Chatterjee, student of M.Sc. 4th semester (Registration No. - 13120911028), Department of Physics, Presidency University, Kolkata, has successfully completed his project work entitled \textit{``Dark Matter Self Interaction and its Impact on Large Scale Structures"} under our supervision for partial fulfilment of the requirements for obtaining the degree of Master of Science in Physics from Presidency University.\\
\\
\vspace{0.7cm}
\begin{flushright}
\vspace{1cm}
\textbf{Prof. Debasish Majumdar}\\
\vspace{1cm}
\textbf{Dr. Suchetana Chatterjee}\\
\end{flushright}
\end{large}
%%%%%%%%%%%%%%%%%%%%%%%%%%%%%%%%%%%%%%%%%%%%%%%%%%%%%%%%%%%%%%%%%%%%%%%%%%%%%%%%%%%%%%%%%%%%%%%%%%
\newpage
\thispagestyle{empty}
\begin{center}
\begin{Large}
\textbf{Declaration of Authorship}\\
\end{Large}
\end{center}
\vspace{0.7cm}
\begin{large}
I, Chayan Chatterjee, hereby declare that this thesis titled \textbf{``Dark Matter Self Interaction and its Impact on Large Scale Structures"} and the work presented in it are my own. This project was carried out by me under the supervision of Prof. Debasish Majumdar and Dr. Suchetana Chatterjee at the Department of Physics, Presidency University, for partial fulfilment of the requirements for obtaining the degree of Master of Science in Physics from Presidency University, Kolkata. I am solely responsible for unintentional oversights and errors in the report, if any. I further declare that it has not formed the basis for the award of any degree, diploma, membership, associateship or similar title of any other university or institution.\\
\\
%\vspace{0.7cm}
\begin{flushright}
\textbf{Chayan Chatterjee}\\
PG-2, Sem-4 (M.Sc.)\\ Regn. No. - 13120911028\\
Department of Physics\\
%Dated : 20.05.2018
\hfill
Presidency University, Kolkata
\end{flushright}
\end{large}
%%%%%%%%%%%%%%%%%%%%%%%%%%%%%%%%%%%%%%%%%%%%%%%%%%%%%%%%%%%%%%%%%%%%%%%%%%%%%%%%%%%%%%%%%%%%%%%%%%
\addcontentsline{toc}{chapter}{Abstract}

\begin{abstract}
The $\Lambda $CDM model of cosmology, though very successful at large scales, has some discrepancy with observations at the galactic and subgalactic scales. These include the `core-cusp problem', `missing satellites problem' etc. Spergel and Steingardt (2000) proposed that if dark matter undergoes feeble self interactions with each other, then such problems can be averted.\\
In this thesis, a two-component dark matter model involving two singlet scalar fields capable of self-interactions has been proposed and its impact on large scale structure formation has been studied through cosmological simulations. The proposed model involves simple extentions of the Standard Model with two singlet scalar fields formed non-thermally through the decay of heavier particles in the very early universe. These particles acquire their relic abundance through a `freeze-in' mechanism. Our model assumes that the FIMP scalar fields have no interaction with any other Standard Model particle except through the Higgs doublet. The coupled Boltzmann equation of the FIMP-FIMP model was solved and the relic densities for different values of the coupling parameters were obtained and matched with PLANCK results. The masses of the FIMP particles were chosen within the allowable range for self-interaction following the prescription of Campbell et.al. (2015) \cite{7}\\
The impact of dark matter self interactions was studied through cosmological simulations  using a modified version of the parallel TreePM code GADGET-2 and the halo mass function and halo catalog for different dark matter self interaction cross sections were obtained. Lastly, the newly developed `Effective Theory of Structure Formation' (ETHOS) framework \cite{55} which is a new and innovative paradigm in the study of the cosmological effects of different dark matter models was studied and using the public code, ETHOS-CAMB \cite{55} the signatures of dark acoustic oscillations in the matter power spectrum for a particular dark matter model was obtained. \\ 
\end{abstract}
%\cleardoublepage

%\addcontentsline{toc}{chapter}{Acknowledgements}

\chapter*{Acknowledgements}

I would like to express my heartfelt gratitude to my advisors, Prof. Debasish Majumdar and Dr. Suchetana Chatterjee for their excellent guidance and support throughout the duration of this project and for instilling in me the necessary self-belief and motivation whenever I needed them. For all of this and much more, I cannot thank them enough.\\
I am extremely grateful to Dr. Gour Bhattacharya, Presidency University, for reading this thesis and providing very valuable suggestions, Dr. Jun Koda, INAF, Italy for some very useful discussions, Dr. Nishikanta Khandai, NISER whose codes I have used in this work.\\
I am also very grateful to my friend Anto Lonappan for teaching me how to install and run GADGET and all the members of Dr. Suchetana Chatterjee's research group `Presi-PACT' for valuable contributions and discussions during the weekly group meetings. \\
My hearfelt gratitude also extends to my classmates Subhendu Saha, Amitava Banerjee, Agniva Roychowdhury and Kanaya Malakar for their help and valuable suggestions.\\
During the course of this thesis work, I had the opportunity to visit the beautiful NISER campus at Bhubaneshwar for the `Introductory School on Galaxy Formation', where besides learning a lot of things that shaped my understanding of the subject, I ran my very first simulation program. I thank the organisers immensely for the experience.

\tableofcontents
%\iffalse

%\body

\chapter{Introduction}
%SC cite planck and some other papers, give reference to all the papers for Bullet and rotational curve etc etc. 
The observational results of PLANCK 2013 \cite{1},\cite{2} collaboration reveal that baryonic (observable) matter constitutes only about 4.8 percent of the total energy budget of the universe. More than 80 percent of the matter content of the universe consists of an unknown, non-luminous, substance, called dark matter whose presence is discerned through it's gravitational interaction with observed matter. Although the particle(s) constituting dark matter is still unknown, several indirect evidences of their existence abound; amongst them being the famous Bullet Cluster \cite{3}, flatness of rotational curves in spiral galaxies \cite{4}, gravitational lensing \cite{5} etc, the details of which have been discussed below. 
\section{Indirect Evidences of Dark Matter}
The first evidence of the existence of dark matter came from the study of velocity dispersion of eight galaxies in the Coma Cluster by Fritz Zwicky (1933). \cite{6} The observed mass of these galaxies were found to be much larger than what can be expected from normal (baryonic) matter. Study of rotation curves of the Andromeda galaxy by Babcock(1939) \cite{8} and of stars in spiral galaxies by Vera Rubin et. al. (1985) \cite{9} showed that the circular velocity becomes constant rather than falling by 1/r$ ^{1/2} $ as expected from Newtonian dynamics. \\
For a star in a galaxy moving with a velocity $ v(r) $ at a distance $ r $ from the centre of the galaxy, the gravitational force balances the centrifugal force, given by:\\
\begin{large}
\begin{equation}
\dfrac{mv(r)^{2}}{r} = \dfrac{GMm}{r^{2}} 
\end{equation}
\end{large}
where, $ M $ is the mass within radius $ r $. Therefore, if the mass within distance $ r $ is constant, which should be the case if there is no dense invisible mass within the centre of galaxy and the star, the celocity should fall off as:\\
\begin{large}
\begin{equation}
v(r) \sim \dfrac{1}{r^{1/2}}
\end{equation}
\end{large}
But observational results show that\\
\begin{large}
\begin{equation}
v(r) \sim constant
\end{equation}
\end{large}
for large $ r $.\\
Therefore, to ensure agreement with Newton's laws, it is believed that all galaxies are sitting inside halos of invisible mass that has no observational signatures and interacts only through gravitational interaction. This invisible mass came to be known as `dark matter'. The rotation curves are shown below in Figure 1.1: \\
% SC give the derivation for rotation curves and why it should be flat. 
\begin{figure}[H]
\centering
  \centering
  \includegraphics[width=10cm]{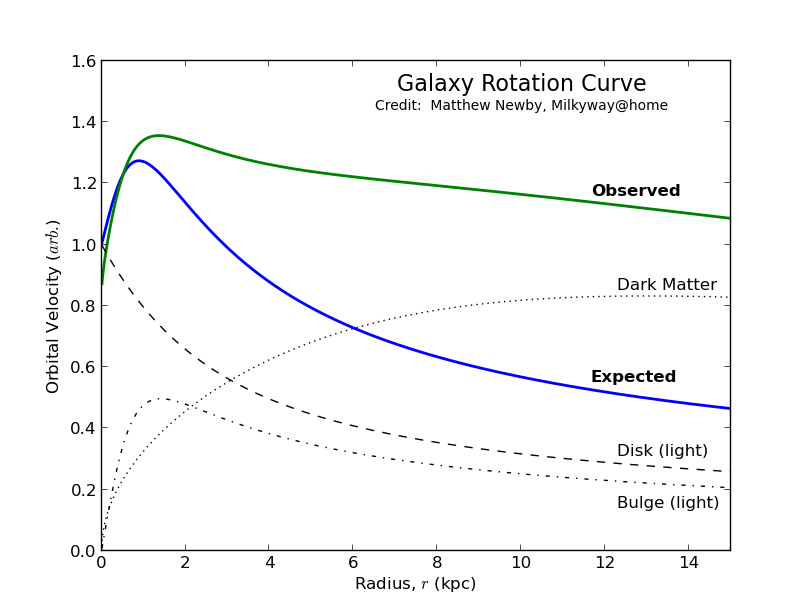}
  \label{fig:1}

\caption{Galaxy rotation curve}
c. Matthew Newby, Milky@home
\end{figure}
% provide some history of other evidences.
\vspace{1cm}
In the intra-cluster medium (ICM), the ambient hot gas appear to be in a state of equilibrium. The hydrostatic pressure of the gas is balanced by the mass of the galaxy.\\
\begin{large}
\begin{equation}
\dfrac{GM(<r)}{r^{2}}4\pi r^{2}\rho_{gas}(r)dr = -4\pi r^{2}\dfrac{dp}{dr}dr
\end{equation}
\end{large} 
where $ \rho_{gas}(r) $ is the density of gas upto radius $ r $, $ p $ is the pressure of the gas and $ M(r) $ is the mass of the galaxy upto radius $ r $.\\
The mass of stars is determined through simple Newtonian two body calculations for binary systems or through the study of main-sequence stars in a Hertzsprung-Russell(HR) diagram which shows an ordering according to star's mass. The luminosity data required for the calculations is obtained by optical satellites. Besides, galaxy clusters have ten times more X-ray gas than stars, as detected by X-ray telescopes like Chandra. The luminosity data of the gas so obtained is thus used to determine it's mass through X-ray luminosity-mass relation of local galaxy clusters \cite{10}. It was concluded from the observations that there is not enough mass in the stars and gas to provide the necessary gravity. In order to satisfy the mass requirement, elliptical galaxies must contain about five times as much mass in dark matter as the amount present in stars and gas.\\ 
One particluar observation strongly hinting at the existence of dark matter is the observation of the Bullet Cluster (Markevitch et al. (2002) \cite{3}). The Bullet Cluster (1E 0657-558) consists of two colliding galaxies. The stars and galaxies constituting the galaxies, being collisionless passed through each other, while the gas, being collisional accumulated in the centre of the colliding masses. Now, if the galaxies consisted of no other form of matter except stars and gas, 90 percent of the total mass of the system is expected to lie at the centre. But gravitational lensing measurements reveal that most of the mass are concentrated at two separated contours lying outside the central region, strongly hinting at the existence of some other invisible form of matter, as evident in Figure 1.2: \\
\begin{figure}[H]
\centering
  \centering
  \includegraphics[width=8cm]{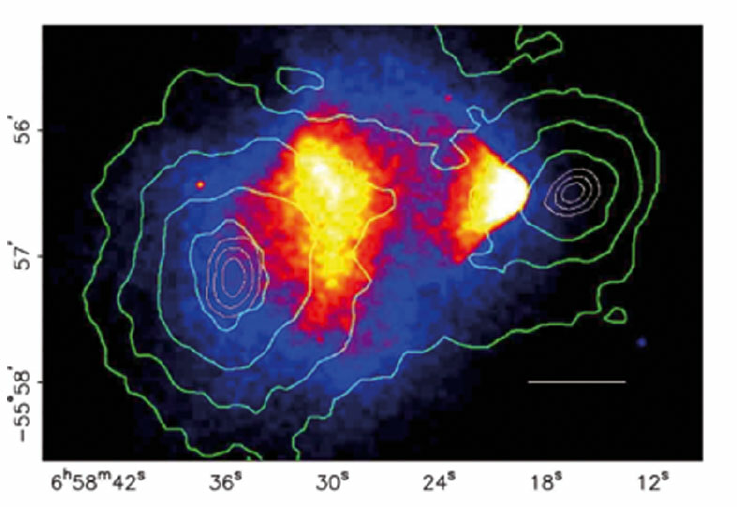}
  \label{fig:1.2}
\caption{Bullet Cluster}
c. NASA/STScI; ESO WFI; Magellan/U.Arizona/ D.Clowe et al. 
\end{figure}
Besides detection via their gravitational influences, there are several other evidences of dark matter. These are called indirect detection. These indirect evidences are typically decay or annihilation products of dark matter particles of different masses. For an example, observations from FermiLAT data have revealed an excess in the $ \gamma $ ray photons from the centre of our galaxy in the energy range 1-3 GeV \cite{4} as we can see in Figure 1.3: \\
\begin{figure}[H]
\centering
  \centering
  \includegraphics[width=9cm]{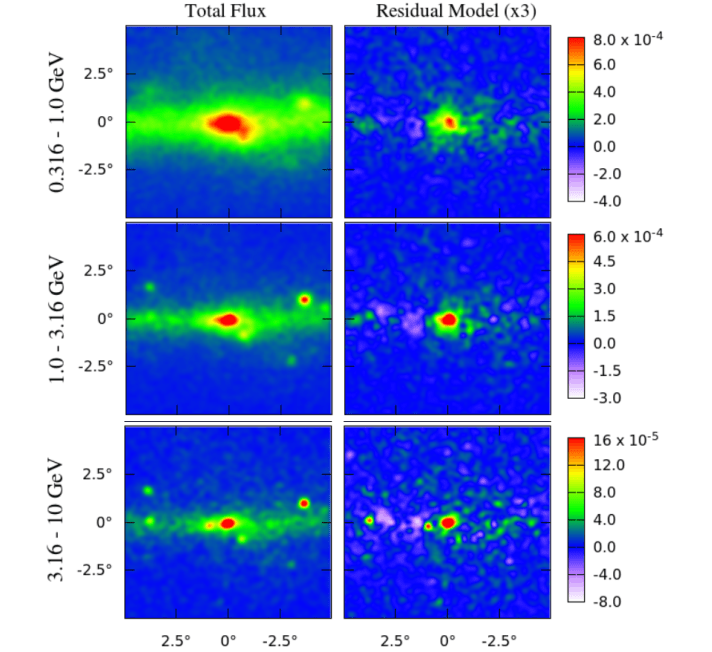}
  \label{fig:1.3}

\caption{Gamma ray count maps of the Galactic Centre region showing a clear                  excess at 1 $\sim $ 3 GeV. c. Daylan, Finkbeiner (2015) \cite{11}}
\end{figure}
Although it was initially speculated that this excess was due to hundreds of milli-second pulsars from the centre of our galaxy \cite{12}, the hypothesis was ruled out, primarily because these pulsars have not been observed and also because the spatial extent of the emission expected from these pulsars is grossly inconsistent with that observed by FermiLAT. Therefore, its origin is considered to be the decay of a dark matter particle, \cite{13},\cite{14},\cite{15} possibly low or intermediate mass, `Weakly Interacting Massive Particle' (WIMP) described later.\\ %SC give reference for this inference
Another indirect evidence of dark matter is the anomolous 3.55 keV X-ray line, observed from the satellite XMM-Newton from an analysis of about 72 different galaxy clusters. \cite{16}. The origin of this line is still unknown, however there are dark matter models which try to explain the same through decay of a very low mass dark matter particle of mass $ \sim $ 7.1 keV. \cite{17},\cite{18},\cite{19}\\
\vspace{1cm}
%\section{Dark Matter detection experiments} 
\section{Dark Matter detection experiments}       
Direct detection experiments aim to detect dark matter particles via their interaction or scattering with atomic nuclei that is kept in a detector buried deep under the earth. This is done to ensure that background neutrons and other particles do not interfere with the experiment. Dark matter being non-interacting with Standard Model particles passes through the layers of the earth and reaches the detector. By studying the change in energy of the detector material during a scattering event, one can calculate the mass and energy of the dark matter particle and therby put tight constraints on particle physics models of dark matter. \cite{20}, \cite{21} %SC (cite papers) 
The signal rate depends on the dark matter mass, local density of dark matter in the region and its velocity distribtion. The necessary dark matter-nucleon scattering cross-section for WIMP dark matter of mass $ \sim $ 100 GeV/c$^{2} $ is smaller than 10$ ^{-42} $ cm$ ^{2} $ \cite{22}
Although no detection event have so far been recorded, direct detection experiments have placed upper limits on the DM-nucleon scattering cross-section, with the tightest constraints coming from experiments like Large Underground Xenon (LUX), Cryogenic Dark Matter Search (CDMS), XENON etc. \\
\textbf{Large Underground Xenon-(LUX)} is a 250 kg xenon experiment located in the Sanford Underground Research facility (SURF), USA. LUX has obtained the strongest reported cross-section limit to date ~ 1.1 $\times$ 10$ ^{-46} $ cm$ ^{2} $ upper limit at a WIMP mass of 50 GeV/c$ ^{2} $.\cite{23} \\ 
The succesor of LUX, called the \textbf{LUX-ZEPLIN} is a 7 ton liquid xenon target, expected to start at 2020. It is projected to reach a sensitivity of 3$ \times $ 10$ ^{-48} $ cm$^{2}$ at 40 GeV/c$^{2}$. \cite{24}
% need citations of all of these
\textbf{XENON100} experiment uses 100 kg liquid xenon as target material. When a direct collision occurs, the liquid scintillates (flashes) because of ionization in the liquid. The lowest cross-section so far obtained is 1.0$\times$ 10$^{-45}$ cm$^{2}$ at 50 GeV/c$^{2}$. \cite{25} \\ 
Other experiments include the \textbf{DarkSide Collaboration} that use liquid argon as the detector material. \cite{26}
Search for lower mass dark matter, of the order of a few GeV/c$ ^{2} $ have been undertaken by the experiments: \textbf{SuperCDMS, CDEX, CoGENT, CRESST}. \cite{27},\cite{28},\cite{29},\cite{30}\\
% cite references
The \textbf{PICO experiment} searches for spin-independent dark matter. \cite{31} \\
The most sensitive dark matter detector in operation, which has been running since 2017 is the \textbf{XENON1T} at Gran Sasso National Laboratory in Italy. It uses pure liquid xenon cooled at -139$ ^{\circ} $F. The projected sensitivity is 2$ \times $ 10$ ^{-47} $ cm$ ^{2} $ at 50 GeV/c$ ^{2} $. \cite{32} \\ 
\begin{figure}[H]
\centering
  \centering
  \includegraphics[width=12cm]{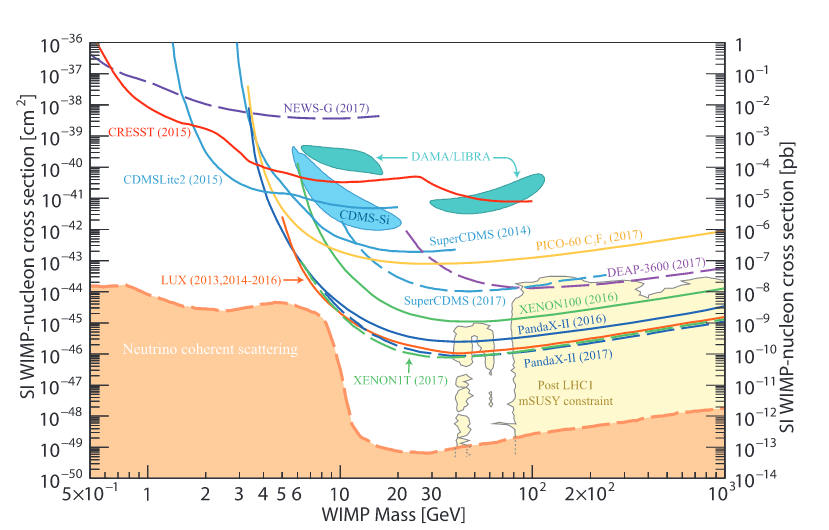}
  \label{fig:1.4}

\caption{WIMP cross sections (normalized to a single nucleon) for spin-independent coupling versus mass. Courtesy: Teresa Marrodán Undagoitia and Ludwig Rauch \cite{69}}
\end{figure}

From Figure 1.4, we can see that SuperCDMS(2014) and PICO (2017) results have found the lower limit of dark matter-nucleon scattering cross-section to be around 10$ ^{-42} $-10$ ^{-45} $ cm$ ^{2} $ for dark matter mass $ \sim $ 7-10 GeV. Theoretical calculations predict such a value of cross-section for WIMP dark matter of mass $ \sim $ 100 GeV. This discrepancy can be explained by considering the existence of low mass dark matter of mass a few GeVs. The recent observation of a dip in the 21 cm signal, at redshifts corresponding to the Cosmic Dawn \cite{70} supports the existence of such low mass WIMPs. This is because the amplitude of the detected dip in the signal is more than twice than expected, indicating that at the epch of Cosmic Dawn, gas must have interacted with something colder, which can only be dark matter. The research suggests that the dark matter particle would not be much heavier than a proton for the dip to be as large as observed, which is well below masses predicted for the WIMP.      

% cite papers for the DAMA experiment and write more details. 
\vspace{1cm}
Other kinds of dark matter detection experiments like DAMA, rely on detecting an annual variation of the number of detection events. DAMA works by detecting flashes of light or scintillations that occur inside crystals of sodium iodide when subatomic particles hit it's nucleus. Such flashes can occur as a result of collisions by stray neutrons and other background sources. But a signal from dark matter in the Milky Way would stand out, because it would show up as a characteristic yearly modulation. \\
This is because as the Sun moves around the Galaxy, the dark matter halo should hit the Solar System as a wind heading in it's diection, which seen from Earth would vary slightly in speed as Earth circles the Sun. Dark-matter detections depend on the speed with which the Earth moves around the Sun. Greater speed implies more detections. So the number of flashes detected would vary throughout the year. The signals should peak in early June and be at their lowest in early December. In the year 1997, DAMA/LIBRA confirmed the presence of a model-independent annual variation in the 2-6 keV range which satisfy all the features of a dark matter signal.\\
However, there exists considerable skepticism in the physics community about the signal being an indicator of dark matter since no direct detection experiment has been able to find any signal for dark matter. Independent experiments like ANAIS \cite{71}, based on the same technique are trying to reproduce the same result. The team will wrap up it's first year of data taking on August 2018. Other experiments like the PICO-LON \cite{72} detector, which Japanese researchers hope to set up in the Kamioka Underground Observatory, will aim to be sensitive to low-energy events. Any detection signal from the experiment can act as a confirmation to the 2010 DAMA results which show signatures of such low-energy collisions. \cite{73}\\ 
\linebreak
In the next chapter, I have described the production mechanisms of dark matter and have described how the relic abundance of dark matter is obtained theoretically.

\chapter{Dark Matter production and Relic Density}
% SC this should go in the beginning. 
\section{Thermal history}
In the early universe, the thermal dark matter particles were in chemical and thermal equilibrium with the hot plasma. The extremely high temperature of the universe in the early epoch caused baryons and dark matter particles to be ultra-relativistic and their very high number density caused them to be in a constant state of collision with each other, establishing the thermal equilibrium. But as the universe expanded, the temperature cooled and the interaction rate of the particles decreased. At a particlular temperature, when the interaction rate of the dark matter particles, $ \Gamma $ became less than the expansion rate of the universe, the dark matter decoupled from the plasma and acquired a relic abundance.
This phenomenon is called `freeze-out'. The comoving density of the particle species therefore became constant. The evolution of such a particle is shown in Figure 2.1. Based on the thermal history, dark matter is classified into two categories: thermal and non-thermal described in detail below.\\

% Give an intro of the thermodynamics of the expanding Universe and explain thermal equilibrium from there. 
\subsection{Thermal dark matter}
Thermal dark matter particles were produced in the very early universe in the post inflation era by collisions of particles in the hot plasma. These particles may have been produced in pairs of particles and anti-particles which would then annihilate to form Standard Model particles. The production and annihilation processes were initially in equilibrium.\\
If the dark matter partice is denoted by $ \chi $ and its number density is denoted by $ n_{\chi} $, then the Boltzmann distribution function gives the evolution of its number density:
\begin{large}
\begin{equation}
 n_{\chi} = n_{\chi} - n_{\bar{\chi}} \sim \left(\frac{m_{\chi}T}{2 \pi}\right) ^{3/2} e^{-m_{\chi}/T}
\end{equation}
\end{large}
Therefore for $ T < m_{\chi} $, the number density falls off exponentially as the tail part of the Boltzmann distribution dominates. This effect is compounded by the cooling effect of the expansion of the universe, which ultimately decreases the interaction rate of particles. Ultimately, when the interaction rate falls below the expansion rate of the universe, freeze-out occurs, and the particles acquire a relic abundance. \cite{34}\\
The relic density is inversely proportional to the annihilation cross section.\\
\begin{large}
\begin{equation}
 \Omega_{\chi} \sim \frac{1}{\langle \sigma v \rangle}
\end{equation}
\end{large} 
where $\Omega_{\chi}$ is the normalised density of the dark matter, $\sigma$ is the annihilation cross-section and $v$ is the relative velocity. The relic density is obtained by solving the following Boltzmann equation:
\begin{large}  
\begin{equation}
\frac{dn_{\chi}}{dt} = -3Hn_{\chi} - \langle \sigma v \rangle \left( n_{\chi}^{2} - {n_{\chi}^{eq}}^{2} \right)
\end{equation}
\end{large}
One popular dark matter candidate produced via thermal mechanism is the Weakly Interacting Massive Particle or WIMP. Their mass can lie in the range between a few GeV and $ \sim $ 100 TeV. It is found that in order to obtain the correct relic abundance of dark matter today via thermal production, the annihilation cross section required is $\langle \sigma v \rangle \simeq$ 3$ \times $ 10$^{-26}$ cm$^{3}$ s$^{-1}$ which matches the calculated weak interaction cross-section value for WIMP dark matter particle of mass $\sim$ 100 GeV, i.e. electroweak scale. This apparent coincidence is called the ``WIMP miracle."

\begin{figure}[H]
\centering
  \centering
  \includegraphics[width=8cm]{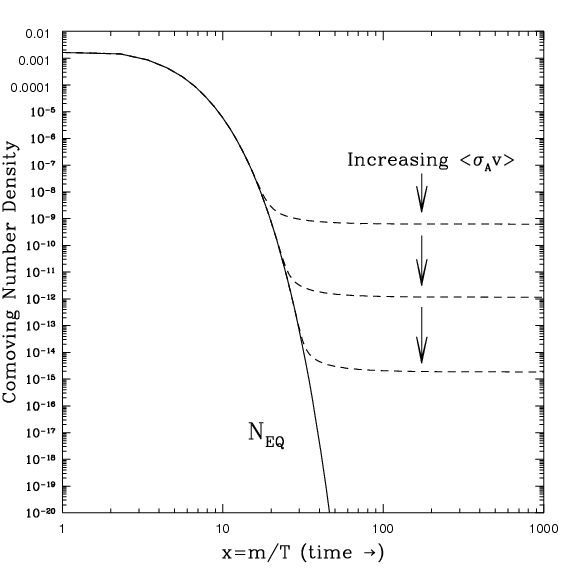}
  \label{fig:3}

\caption{Thermal freeze-out of dark matter for different annihilation cross-sections}
c. Dan Hooper, [hep-ph] FERMILAB-CONF-09-025-A 
\end{figure}

\subsection{Non thermal dark matter}
% SC give references for non-thermal dark matter
Unlike thermal dark matter, non-thermal dark matter was never in thermal and chemical equilibrium with the hot plasma in the early universe. This is because the interaction strength of these dark matter particles were extremely less, much less than those produced by the thermal paradigm. These type of particles were produced mainly from the decay of heavier particles in the early universe. They might also have been produced by the annihilation of particles in the hot plasma, but the rate of occurence of such a process is extremely less compared to the production from decay of heavier particles. In this scenario, the dark matter relic density is generated from a different mechanism, known as `Freeze-in', which is exactly opposite to the freeze-out process. \cite{35} 
%SC explain the freeze in 
Non-thermal dark matter is therefore slowly produced through annihilations of SM particles, until the universe becomes too cool for dark matter to be produced. The resulting dark matter abundance thus slowly increases or `freezes in' from zero total density to the present value, never actually the equilibrium value.\\

Dark matter particles produced through this mechanism is generally called `Feebly Interacting Massive Particles' or FIMP and they have masses smaller than thermally produced WIMPs. Some popular candidates for FIMPs are winos and axions, which may have been produced by a non-thermal mechanism. \cite{36} The evolution of number density of non-thermal dark matter is depicted in Figure 2.2:\\
 
\begin{figure}[H]
\centering
  \centering
  \includegraphics[width=10cm]{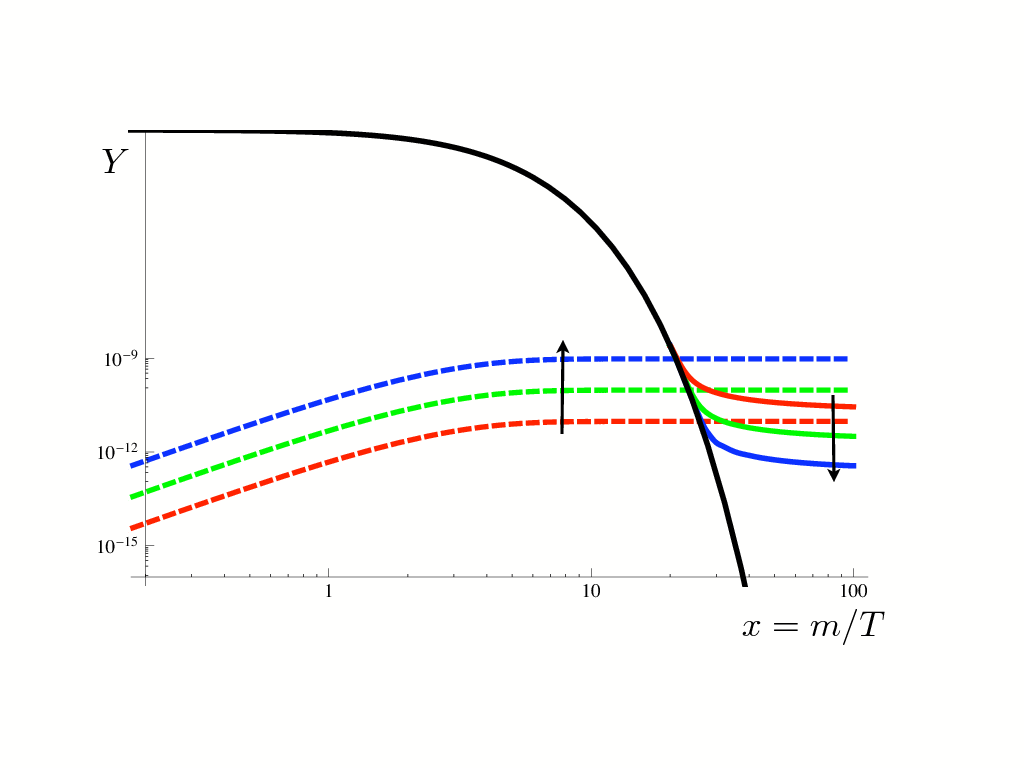}
  \label{fig:3}

\caption{Non-thermal freeze in}
c. Hall et. al. (2009)
\end{figure}
The black solid line indicates dark matter density if equilibrium is maintained. The solid coloured lines indicate freeze-out mechanishm for increasing coupling strengths, denoted by arrows and the dotted coloured lines indicate freeze-in mechanism.\\

\chapter{Dark Matter Self Interactions}

\section{Problems with $ \Lambda $CDM model}
The cosmological concordence model, called the $ \Lambda $CDM, which describes dark matter as 'cold' (non-relativistc) and collisionless has been very successful in describing several large scale cosmological phenomenon like CMB fluctuations, structure formation and primordial abundances \cite{37}, but has been observed to fail at galactic and sub-galactic scales, as suggested by N-body simulations performed with cold dark matter since the 1990s.\cite{38} Some of these problems have been discussed below:\\
% SC cite cosmology papers 

\subsection{The Core-Cusp Problem}

\begin{figure}[h]
\centering
  \centering
  \includegraphics[width=8cm]{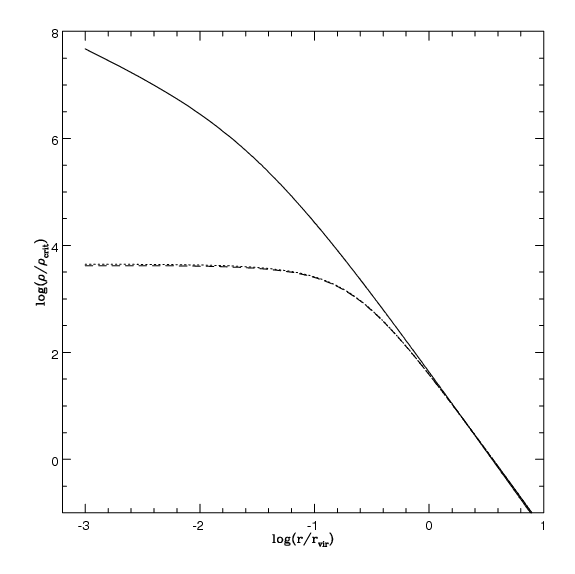}
  \label{fig:3}

\caption{Core vs. Cusp}
c. Del Popolo, Astrophys.J.698(2009)
\end{figure}

Collisionless CDM simulations by Navarro, Frenk and White \cite{39} and many other groups hence, predict a steep density profile  with $ \rho \propto \frac{1}{r} $ in the central regions.\\
\begin{large}  
\begin{equation}
\rho_{NFW} = \dfrac{\rho_{s}}{\left(r/r_{s}\right) \left(1+r/r_{s}\right)^{2}}
\end{equation}
\end{large}
However, Flores and Primack \cite{40} ruled out cuspy profiles from DDO galaxy rotation curves, and showed them to be well approximated by cored (or pseudo-) isothermal density profiles. This is known as the core-cusp problem. The steep central density in CDM simuations is called a `cusp' and the flat density profile is called a `core'.
%SC cite simulations and observational papers

\subsection{Too big to fail problem}
The Milky Way's brightest dSphs (dwarf spheroidal galaxies) are expected to lie in the most massive sub halos. But CDM only simulations predict too high central densities to host the observed satellites. It is called `too big to fail problem' because of the assumption that the Milky Way is `too big' to fail to form big stars.\cite{41}
%SC SC Cite papers
\subsection{Missing satellites problem} 
Cosmological simulations suggest that a Milly Way sized halo should have around 500 subhalos around it, capable of hosting dwarf galaxies and satellites. \cite{42} But so far only about 50 such satellites have been discovered. \cite{43}\\
 The MW has bright dSphs, Sagittarius, the LMC and the SMC, thus much less than the
500 satellites, obtained in simulations, with larger circular velocities than Draco and Ursa-Minor. \cite{41}
%SC cite simulations and observational papers. 
\section{Why dark matter self interactions?}
It has been proposed by Spergel and Steinhardt in the year 2000 \cite{46} that these small-scale problems of $ \Lambda $CDM can be alleviated by self-interacting dark matter, with a self -scattering cross-section $ \sigma $  over DM mass, $ m $ in the range 0.1 $ \leq $ $ \sigma / m $ $ \leq $ 10 cm$^{2}$/g \cite{44}. 
It is believed that self-scattering of DM particles lead to a heat transfer that decrease the density in the centre of DM halos, thereby turning cuspy profiles into a cored profile and also reduces the subhalo population because of the reduced central density. Therefore, self-interacting dark matter addresses two major small-scale problems of LCDM - the core-cusp and the missing satellites problem. These results have been confirmed by several groups like Yoshida et. al. (2000) \cite{45} and others for DM-only simulations with different self-interaction cross-sections. Below are the dark matter halo density profiles and subhalo population plots obtained by simulations of Yoshida's group:
\begin{figure}[H]
\centering
\begin{subfigure}{.5\textwidth}
  \centering
  \includegraphics[width=8cm]{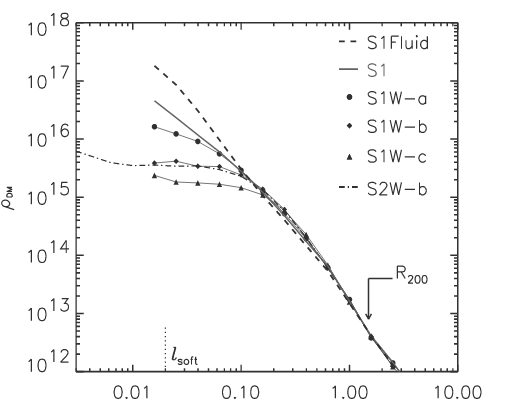}
  \caption{}
  \label{fig:1}
\end{subfigure}%
\begin{subfigure}{.5\textwidth}
  \centering
  \includegraphics[width=9.5cm]{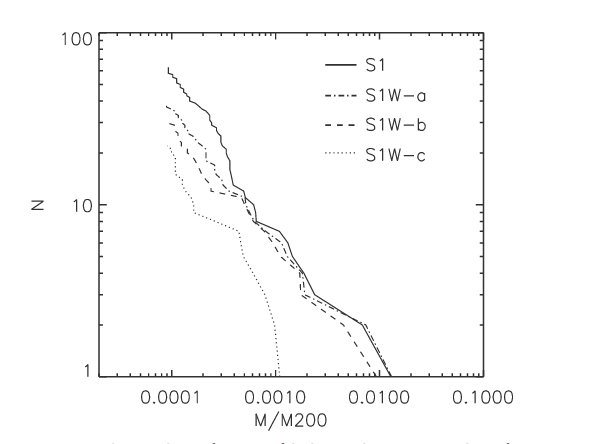}
  \caption{}
  \label{fig:2}
\end{subfigure}
\caption{(a)Change in density profiles of DM halos with increasing self interaction cross-section (b)The total number of subhalos within R$_{200}$ plotted as a function of it's mass in units of M$_{200}$}
\end{figure}

\subsection{Constraints on self interaction cross sections}
Several galaxy cluster mergers have been analysed in order to constrain this self-interaction cross-section. The absence of an offset between the stars and the DM halo in the Bullet Cluster constrains $ \sigma / m $ to $ \leq $ 1.25 cm$^{2}$/g at 68 \% CL. \cite{3} Observations of galaxy cluster Abell 3827, consisting of four central galaxies reveal an offset between the stars and the DM halo in one of the four systems, suggesting a $ \sigma / m $ $ \sim $ 1.5 cm$^{2} $/g \cite{47}. Similar observations of cluster collisions constrain $ \sigma / m $ $ \leq $ 0.47 cm $^{2} $/g. \cite{48}\\

In general, astrophysical objects of different masses are affected differently by DM self interactions. Kaplinghat et al. 2015 \cite{49} found a value of $ \sim $ 2 cm$^{2} $/g consistent with observations of dwarf and low surface brightness galaxies while a value of $ \sim $ 0.1 cm$ ^{2} $/g is necessary to explain observations of density profiles of galaxy clusters. Therefore, velocity dependent cross-sections are also invoked to explain the different observations at different length and mass scales.

\subsection{Conditions for self-interaction}
From the observation of spatial offsets between dark matter halo and stars in galaxies in the Abell 3827 cluster, the dark matter self interaction cross-section has been constrained to $ \sigma_{DM}/m $= 1-1.5 cm$ ^{2} $/g. Using this information, Campbell et. al.\cite{7} have constrained the single scalar dark matter model coupled to the Standard Model and two Higgs doublet models. They have reported that a light dark matter of mass less than 0.1 GeV and produced by the freeze-in mechanism can have the necessary cross-section for self interaction. The analytical expression for self interaction cross section is given below \cite{7}\\
\begin{large}
\begin{equation}
\sigma_{DM}/m = \frac{9 \lambda_{eff}^{2}}{2 \pi m^{3}}
\end{equation}
\end{large}   
Here, $ m $ is the mass of the dark matter particle and $ \lambda_{eff} $ is the quartic coupling term of the dark matter field. The quartic couplic term should be less than $ 2 \pi / 3 $  to satisfy perturbative unitarity bound which arises from the requirement that the S-matrix of all interaction processes must be unitary. $ \left(S^{\dagger} S =1 \right) $. Using equation (2.1) I have obtained the parameter space for mass and quartic coupling for scalar dark matter. The figure obtained is shown below:\\
\begin{figure}[H]
\centering
  \centering
  \includegraphics[width=8cm]{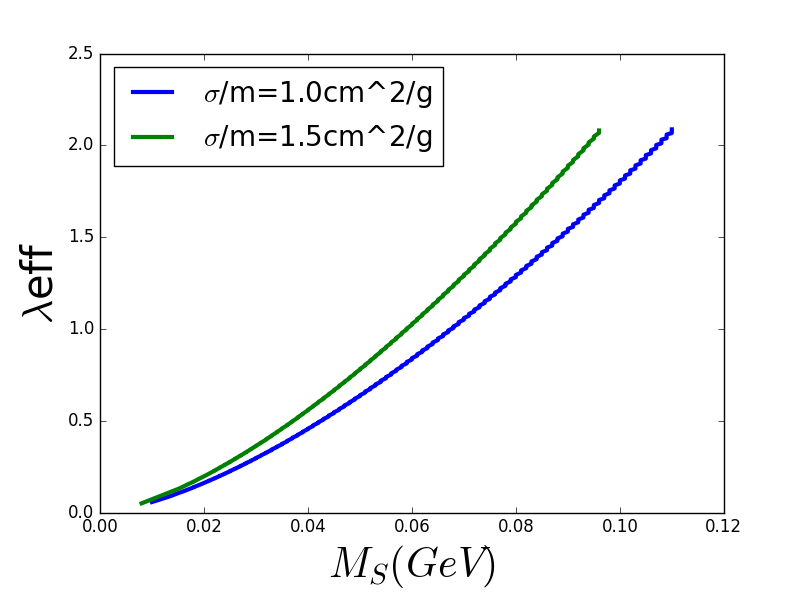}
  \label{fig:3}
\caption{Quartic coupling vs. Mass}
\end{figure}

In this project, I have worked with two two-component singlet scalar dark matter models, where one of the dark matter components is capable of undergoing self-interactions, following the prescriptions of Campbell et.al. (2015). In the next two chapters, I shall describe the two models and the analytical and numerical calculations involved in obtaining the relic density of the dark matter components of these models in keeping with the relic density limits set by PLANCK data. 

Next, I shall describe in subsequent chapters the results of cosmological simulations obtained using the parallel TreePM code, GADGET-2 and its modified version for self-interacting dark matter, the SIDM-GADGET, \cite{50} developed by Dr. Jun Koda, INAF, Italy and the analysis which I have done using those results.
\chapter{WIMP-FIMP Dark Matter model }

In this model, the Standard Model of particle physics is extended by two scalar fields $ \phi_{1} $ and $ \phi_{2} $, both of which are singlets under the Standard Model gauge group SU(2)$ _{L} $ $\times $ U(1)$_{Y}$. The $ \phi_{1} $ field is assumed to be a thermally produced WIMP dark matter candidate and the $ \phi_{2} $ field is a non-thermally produced FIMP dark matter scalar field. These fields couple among themselves but have no other interaction with Standard Model particles except through the Higgs doublet, $ H $. 

\section{Description of the model}
The Lagrangian of the model can be written as:\\
\begin{large}
\begin{equation}
L=L_{SM}+L_{DM}+L_{int} 
\end{equation}
\end{large}
where $L_{SM}$, $L_{DM}$, $L_{int}$ are the terms in the lagrangian corresponding to Standard Model fields, dark matter fields, $ \phi_{1} $ and $ \phi_{2} $ and the inetraction terms of the dark matter fields respectively.\\
The terms corresponding to fields $ \phi_{1} $ and $ \phi_{2} $ are :\\
\begin{large}
\begin{equation}
 L_{\phi_{1}}=\dfrac{1}{2}\partial_{\mu}\phi_{1}\partial^{\mu}\phi_{1}-\dfrac{\mu_{\phi_{1}^{2}}}{2}\phi_{1}^{2}-\dfrac{\lambda_{\phi_{1}}}{4}\phi_{1}^{4}
\end{equation}
\end{large}

\begin{large}
\begin{equation}
L_{\phi_{2}} = \dfrac{1}{2} \partial_{\mu} \phi_{2} \partial^{\mu} \phi_{2}- \dfrac{\mu_{\phi_{2}^{2}}}{2} \phi_{2}^{2}- \dfrac{\lambda_{\phi_{2}}}{4} \phi_{2}^{4} 
\end{equation}
\end{large}

In order to ensure that the scalar fields  $ \phi_{1} $ and $ \phi_{2} $ are stable and have no interaction with Standard Model fermions, a $ Z_{2} \times Z_{2}' $ symmetry is imposed under which either of the fields $ \phi_{1} $ and $ \phi_{2} $ are odd. This symmetry implies that if the sign of the fields $ \phi_{1} $ and $ \phi_{2} $ are changed, i.e., $\phi_{1} \rightarrow - \phi_{1}$ and $\phi_{2} \rightarrow - \phi_{2}$, the Lagrangian itself should remain invariant. ($ L \rightarrow L $). Therefore all terms in the potential containing an odd number of fields will become zero, in order to satisfy this criteria. The resultant potential is the following:\\
\begin{large}
$ V(H,\phi_{1},\phi_{2})= \mu_{H}^{2}H^{\dagger} H+\lambda_{H}(H^{\dagger}H)^{2}+\dfrac{\mu_{\phi_{1}}^{2}}{2}\phi_{1}^{2}+\dfrac{\lambda_{\phi_{1}}}{4}\phi_{1}^{4}+\dfrac{\mu_{\phi_{2}^{2}}}{2}\phi_{2}^{2}+\dfrac{\lambda_{\phi_{2}}}{4}\phi_{2}^{4}+\lambda_{H\phi_{1}}H^{\dagger}H\phi_{1}^{2}+\lambda_{H\phi_{2}}H^{\dagger}H\phi_{2}^{2}+ \lambda_{\phi_{1}\phi_{2}}\phi_{1}^{2}\phi_{2}^{2}+\alpha\phi_{1}\phi_{2}$ \\
\end{large}

Here, $ H $ is the Standard Model Higgs doublet given by: \\
\begin{center}
\(
H=\dfrac{1}{\sqrt{2}}\begin{bmatrix}
0\\\nu+h

\end{bmatrix}
\)

\end{center}
where, $\nu $ is the vacuum expectation value (VEV) of Higgs with a value ~ 246 GeV and $ h $ is the physical Higgs field.\\ 
The $\phi_{2}$ field has a VEV $ \nu_{1} $ whose value lies in the range 2 MeV $ \leq u \leq$ 10 MeV. \\
The $ Z_{2} \times Z_{2}' $ symmetry is softly broken to a residual $ Z_{2}'' $ symmetry under which both $ \phi_{1} $ and $ \phi_{2} $ are odd. This is done by explicitly adding the term $ \alpha\phi_{1}\phi_{2}$ in the potential $V$. It is added in order to ensure diagonalizability of the mass-squared matrix which is obtained by minimising the potential with respect to the fields.\\

\section{Obtaining the masses}
Now in order to obtain the mass of the fields $ h $, $ \phi_{1} $ and $ \phi_{2} $, the potential is minimised with respect to the fields, considering all other fields to be zero.\\
In order to do that, first the physical value of the fields plus their VEV, i.e., \(
H=\dfrac{1}{\sqrt{2}}\begin{bmatrix}
0\\\nu+h

\end{bmatrix}
\)
$ \phi_{1}=\phi_{1} $ and $\phi_{2} = \phi_{2}+ \nu_{1}$ are substituted into the expression of the potential and then the partial derivatives are calculated
\begin{center}
\begin{equation}
 \left(\dfrac{\partial V}{\partial h}\right)_{h=0,\phi_{1}=0,\phi_{2}=0}=\mu_{H}^{2}+\lambda_{H}\nu^{2}+\lambda_{H\phi_{1}}\nu_{1}^{2}=0
\end{equation}
\begin{equation}
 \left(\dfrac{\partial V}{\partial \phi_{1}}\right)_{h=0,\phi_{1}=0,\phi_{2}=0}=\mu_{\phi_{1}}^{2}+\lambda_{\phi_{1}}\nu_{1}^{2}+\lambda_{H\phi_{1}}\nu^{2}=0 
 \end{equation}
\hspace{120pt}
\begin{equation}
\left(\dfrac{\partial V}{\partial \phi_{2}}\right)_{h=0,\phi_{1}=0,\phi_{2}=0}=\alpha\nu_{1}=0 
\end{equation}
\end{center}
By calculating these derivatives we get three relations, which when substituted into the expressions of the double derivative of the potential with respect to each of the fields, the mass-squared matrix is obtained:
\begin{center}
\begin{large}

\(
M_{scalar}^{2}=
  \begin{pmatrix}
    2\lambda_{H}\nu^{2} & 2\lambda_{H\phi}\nu_{1}\nu & 0 \\
    2\lambda_{H\phi_{1}}\nu_{1}\nu & 2\lambda_{\phi_{1}}\nu_{1}^{2} & \alpha \\
    0& \alpha & \mu_{\phi_{2}^{2}+\lambda_{H\phi_{2}}\nu^{2}+2\lambda_{\phi_{1}\phi_{2}}\nu_{1}^{2}}
  \end{pmatrix}
\)

\end{large}
\end{center}

In order to obtain the eigenvalues, the matrix needs to be diagonalised by a unitary transformation, for which the  Pontecorvo–Maki–Nakagawa–Sakata (PMNS) matrix is used:\\
\begin{center}
\begin{large}

\(
U=
  \begin{pmatrix}
    c_{13}c_{12} & s_{12}c_{13} & s_{13} \\
    -s_{12}c_{23}-s_{23}s_{13}c_{12} & c_{23}c_{12}-s_{23}s_{13}s_{12} & s_{23}c_{13} \\
    s_{23}s_{12}-s_{13}c_{23}c_{12}& -s_{23}c_{12}-s_{13}s_{12}c_{23} & c_{23}c_{13}
  \end{pmatrix}
\)

\end{large}
\end{center}
Here, $ c_{ij} $ and $ s_{ij} $ represent cos $ \theta_{ij} $ and sin $ \theta_{ij} $ respectively, where i,j runs from 1 to 3. The angles $ \theta_{12} $, $ \theta_{23} $ and $ \theta_{13} $ are rotation angles needed to diagonalise the mass-squared matrix. The values of these angles are assumed to be very small.

The eignevalues obtained are denoted by $ M_{\chi_{1}}, M_{\chi_{2}}, M_{\chi_{3}}  $, whose masses are: $ M_{\chi_{1}} $ ~ 125 GeV, $ M_{\chi_{2}} \simeq$ 80-110 GeV and $ M_{\chi_{3}} \sim$ 7 keV. This mass range was chosen from the motivation that the heavier WIMP dark matter particle would explain the FermiLAT gamma-ray excess and the lighter FIMP particle would explain the 3.55 keV X-ray line observed from XMM Newton data as well as undergo self-interactions.

\section{The Boltzmann equations}
The relic abundance of the individual dark matter candidates are obtained by simulataneously solving the coupled Boltzmann equation of the particular dark matter model using a numerical code. For this work, I have modified the prescription of \cite{51} based on a different model to accomodate the WIMP-FIMP dark matter model.\\
% SC If you have written the Boltzmann code that is great if not cite the code from some previous paper
The coupled Boltzmann equation of the WIMP-FIMP model is the following \cite{60}: \\

\begin{center}
$\dfrac{dY_{\chi_{2}}}{dz}= -\left( \dfrac{45G}{\pi}\right) ^{-\frac{1}{2}}\dfrac{m_{\chi_{2}}}{z^{2}}g_{*}^{\frac{1}{2}}\left[ \left\langle \sigma v\right\rangle _{\chi_{2}\chi_{2}\rightarrow x\bar{x}}\left(Y_{\chi_{2}}^{2}-\left(Y_{\chi_{2}}^{eq}\right) ^{2}\right)+\left\langle \sigma v\right\rangle _{\chi_{2}\chi_{2}\rightarrow h_{3}h_{3}}Y_{\chi_{2}}^{2}\right]$  
\end{center}
     
\begin{center}
$\dfrac{dY_{\chi_{3}}}{dz}= -\dfrac{2M_{Pl}}{1.66M_{\chi_{1}}^{2}}\dfrac{z\sqrt{g_{*}(T)}}{g_{s}(T)}\Big( \left\langle \Gamma_{\chi_{1}\rightarrow \chi_{3}\chi_{3}}\right\rangle \left( Y_{\chi_{3}}-Y^{eq}_{\chi_{3}}\right) \Big) -\dfrac{4\pi^{2}}{45}\dfrac{M_{Pl}M_{\chi_{1}}}{1.66}\dfrac{\sqrt{g_{*}(T)}}{z^{2}}\times \Big( \Sigma \left\langle \sigma v_{x \bar{x} \rightarrow \chi_{3}\chi_{3}} \right\rangle \left( Y_{\chi_{3}}^{2}-(Y^{eq}_{\chi_{3}})^{2}\right) -\left\langle \sigma v_{\chi_{2}\chi_{2} \rightarrow \chi_{3}\chi_{3}}\right\rangle Y_{\chi_{2}}^{2} \Big)  $
\end{center}
where,  x=fermions, W-bosons, Z-bosons, $\chi_{2},\chi_{3} $\\
Here, $ Y_{\chi_{2}}=\frac{n_{\chi_{2}}}{s} $ and $ Y_{\chi_{3}}=\frac{n_{\chi_{3}}}{s} $ are the comoving number densities of $ \chi_{2} $ and $ \chi_{3} $ respectively and $ z = \frac{M_{\chi_{1}}}{T} $, where $ T $ is the photon temperature and $ s $ si the entropy density of the universe. $ M_{Pl} $ is the Planck mass and $ g_{*} $ is given by:\\
\begin{center}
$ \sqrt{g_{*}(T)}=\dfrac{g_{s}(T)}{\sqrt{g_{\rho}(T)}}\Big(1+\dfrac{1}{3}\dfrac{dln g_{s}(T)}{dln T} \Big)$
\end{center}
$ g_{\rho}(T) $ and $ g_{s}(T) $ are the effective degrees of freedom of energy and entropy respectively. Different processes have different values of decay widths $ \Gamma $ and annihilation cross section $\langle \sigma v\rangle $. Their expressions are given by:\\
\begin{center}
$ \langle \Gamma_{\chi_{1}\rightarrow \chi_{j}\chi_{j}}\rangle = \Gamma_{\chi_{1}\rightarrow \chi_{j}\chi_{j}}\dfrac{K_{1}(z)}{K_{2}(z)}$,\hspace{30pt} j=2,3.\\
\vspace{10pt}
$ \langle \sigma v_{x\bar{x}\rightarrow \chi_{j}\chi_{j}} \rangle = \dfrac{1}{8M_{x}^{4}TK_{2}^{2}(\frac{M_{x}}{T})}\bigint_{4M_{x}^{2}}^{\infty} \sigma_{xx \rightarrow \chi_{j}\chi_{j}}(s - 4M_{x}^{2})\sqrt{s}K_{1}\dfrac{\sqrt{s}}{T} ds$,\\
\end{center}
In the above equations, $ K_{1} $ and $ K_{2} $ are modified Bessel functions of first and second order respectively and $ s $ is the Mandelstam variable, which is equal to the square of the total momentum at each vertex of a Feynman diagram.

In order to solve the Boltzmann equation the expressions of the decay widths and annihilation cross-sections of each of the production and decay channels are incorporated into the code and is solved to obtain the relic density.\\
The motivation behind studying the WIMP-FIMP model was to explore a two-component dark matter model that encapsulates both thermal and non-thermal dark matter physics as well as is able to explain observations of possible dark matter signatures characteristic of both the regimes. The full spectrum of theoretical possibilities arising from a scalar WIMP-FIMP dark matter model can be explored later. For this work, however I have focussed on the FIMP-FIMP model, mainly from the motivation that WIMPs have in recent years, fallen out of favour as a viable dark matter candidate since more than 30 years of direct detection efforts have yielded null results.\\  
I have solved the numerical Boltzmann code and obtained the relic density for different values of the quartic coupling parameters for the FIMP-FIMP model, which is described in the next chapter. 

\chapter{FIMP-FIMP Dark Matter model }

In this model, the Standard Model is extended by two scalar fields $ S_{2} $ and $ S_{3} $, both of which are singlets under the Standard Model gauge group SU(2)$ _{L} $ $\times $ U(1)$_{Y}$. Here the two dark matter components are assumed to be produced via a non-thermal mechanism because of their extremely feeble coupling with other particles. Just like the previous model it is assumed that the fields couple among themselves but have no other interaction with Standard Model particles except through the Higgs doublet, $ H $. This model is described in detail in Biswas et.al.(2015) \cite{51}.\\

\section{Description of the model}
The Lagrangian of the model is given by:\\
\begin{center}
$ \mathcal{L}= \Big(D_{\mu}H\Big)^{\dagger}\Big(D^{\mu}H\Big)+\dfrac{1}{2}\partial_{\mu}S_{2}\partial^{\mu}S_{2}+\dfrac{1}{2}\partial_{\mu}S_{3}\partial^{\mu}S_{3}-V(H,S_{2},S_{3}) $
\end{center}

A $ Z_{2} \times Z_{2}' $ symmetry softly broken to a residul $ Z_{2}'' $ symmetry is imposed on the two fields to ensure their stability just like in the WIMP-FIMP model.  The resultant potential is:\\

\begin{center}

$ V(H,S_{2},S_{3})=\kappa_{1}\Big(H^{\dagger}H-\dfrac{v^{2}}{2}\Big)^{2}+\dfrac{\kappa_{2}}{4}S_{2}^{4}+\dfrac{\kappa_{3}}{4}(S_{3}^{2}-u^{2})^{2}+\dfrac{\rho_{2}^{2}}{2}S_{2}^{2}+\lambda_{12}\Big(H^{\dagger}H\Big)S_{2}^{2}+\lambda_{23}S_{2}^{2}S_{3}^{2}+ \lambda_{13}\Big(H^{\dagger}H - \dfrac{v^{2}}{2}\Big)(S_{3}^{2}-u^{2}) $

\end{center}

where $ v $ is the VEV of Higgs and has a value $ \sim $ 246 GeV and $ u $ is the VEV of $ S_{3} $ lying in the range 2 MeV $ \leq u \leq$ 10 MeV. $ S_{2} $ does not have a VEV.\\

The term $ \alpha S_{2}S_{3}$ in the potential $V$ is explicitly added as a soft breaking term.

\section{Obtaining the masses}
The method of deriving the mass-squared matrix has been explained in the previous chapter. Following the same procedure, the matrix obtained is:\\

\begin{center}
\begin{large}
\(
M_{scalar}^{2}=
  \begin{pmatrix}
    2\kappa_{1}v^{2} & 0 & 2\lambda_{13}uv \\
    0 & \rho_{2}^{2}+\lambda_{12}v^{2}+2\lambda_{23}u^{2} & \alpha \\
    2\lambda_{13}uv& \alpha & 2\kappa_{3}u^{2}
  \end{pmatrix}
\)

\end{large}
\end{center}

The eignevalues obtained from this matrix after diagonalisation using the regular PMNS matrix are $ M_{\chi_{1}}, M_{\chi_{2}}, M_{\chi_{3}}  $, whose masses are: $ M_{\chi_{1}} $ $ \sim $ 125 GeV, $ M_{\chi_{2}} \simeq$ 50-80 GeV and $ M_{\chi_{3}} \sim$ 7 keV-0.1 GeV\\ 

\section{Calculation of relic density}
The relic abundance of the individual dark matter candidates were obtained by simulataneously solving the coupled Boltzmann equation of the particular dark matter model using the prescription in Biswas et.al.\cite{12}\\
The coupled Boltzmann equation of the FIMP-FIMP model is the following: \\

\begin{center}
$\dfrac{dY_{\chi_{2}}}{dz}= -\dfrac{2M_{Pl}}{1.66M_{\chi_{1}}^{2}}\dfrac{z\sqrt{g_{*}(T)}}{g_{s}(T)}\Big( \left\langle \Gamma_{\chi_{1}\rightarrow \chi_{2}\chi_{2}}\right\rangle \left( Y_{\chi_{2}}-Y^{eq}_{\chi_{1}}\right) \Big) -\dfrac{4\pi^{2}}{45}\dfrac{M_{Pl}M_{\chi_{1}}}{1.66}\dfrac{\sqrt{g_{*}(T)}}{z^{2}}\times \Big( \Sigma \left\langle \sigma v_{x \bar{x} \rightarrow \chi_{2}\chi_{2}} \right\rangle \left( Y_{\chi_{2}}^{2}-(Y^{eq}_{\chi_{1}})^{2}\right) -\left\langle \sigma v_{\chi_{2}\chi_{2} \rightarrow \chi_{3}\chi_{3}}\right\rangle Y_{\chi_{2}}^{2} \Big)  $
\end{center}
     
\begin{center}
$\dfrac{dY_{\chi_{3}}}{dz}= -\dfrac{2M_{Pl}}{1.66M_{\chi_{1}}^{2}}\dfrac{z\sqrt{g_{*}(T)}}{g_{s}(T)}\Big( \left\langle \Gamma_{\chi_{1}\rightarrow \chi_{3}\chi_{3}}\right\rangle \left( Y_{\chi_{3}}-Y^{eq}_{\chi_{1}}\right) \Big) -\dfrac{4\pi^{2}}{45}\dfrac{M_{Pl}M_{\chi_{1}}}{1.66}\dfrac{\sqrt{g_{*}(T)}}{z^{2}}\times \Big( \Sigma \left\langle \sigma v_{x \bar{x} \rightarrow \chi_{3}\chi_{3}} \right\rangle \left( Y_{\chi_{3}}^{2}-(Y^{eq}_{\chi_{3}})^{2}\right) -\left\langle \sigma v_{\chi_{2}\chi_{2} \rightarrow \chi_{3}\chi_{3}}\right\rangle Y_{\chi_{2}}^{2} \Big)  $
\end{center}
where,  x=fermions, W-bosons, Z-bosons, $\chi_{2},\chi_{3} $\\
Here, $ Y_{\chi_{2}}=\frac{n_{\chi_{2}}}{s} $ and $ Y_{\chi_{3}}=\frac{n_{\chi_{3}}}{s} $ are the comoving number densities of $ \chi_{2} $ and $ \chi_{3} $ respectively and $ z = \frac{M_{\chi_{1}}}{T} $, where $ T $ is the photon temperature and $ s $ si the entropy density of the universe. $ M_{Pl} $ is the Planck mass and $ g_{*} $ is given by:\\
\begin{center}
$ \sqrt{g_{*}(T)}=\dfrac{g_{s}(T)}{\sqrt{g_{\rho}(T)}}\Big(1+\dfrac{1}{3}\dfrac{dln g_{s}(T)}{dln T} \Big)$
\end{center}
$ g_{\rho}(T) $ and $ g_{s}(T) $ are the effective degrees of freedom of energy and entropy respectively. Different processes have different values of decay widths $ \Gamma $ and annihilation cross section $\langle \sigma v\rangle $. Their expressions are given by:\\
\begin{center}
$ \langle \Gamma_{\chi_{1}\rightarrow \chi_{j}\chi_{j}}\rangle = \Gamma_{\chi_{1}\rightarrow \chi_{j}\chi_{j}}\dfrac{K_{1}(z)}{K_{2}(z)}$,\hspace{30pt} j=2,3.\\
\vspace{10pt}
$ \langle \sigma v_{x\bar{x}\rightarrow \chi_{j}\chi_{j}} \rangle = \dfrac{1}{8M_{x}^{4}TK_{2}^{2}(\frac{M_{x}}{T})}\bigint_{4M_{x}^{2}}^{\infty} \sigma_{xx \rightarrow \chi_{j}\chi_{j}}(s - 4M_{x}^{2})\sqrt{s}K_{1}\dfrac{\sqrt{s}}{T} ds$,\\
\end{center}
In the above equations, $ K_{1} $ and $ K_{2} $ are modified Bessel functions of first and second order respectively and $ s $ is the Mandelstam variable.\\
The Feynman diagrams corresponding to the allowed production and decay channels are shown \cite{51}:\\ 
\begin{figure}[H]
\centering
  \centering
  \includegraphics[width=15cm]{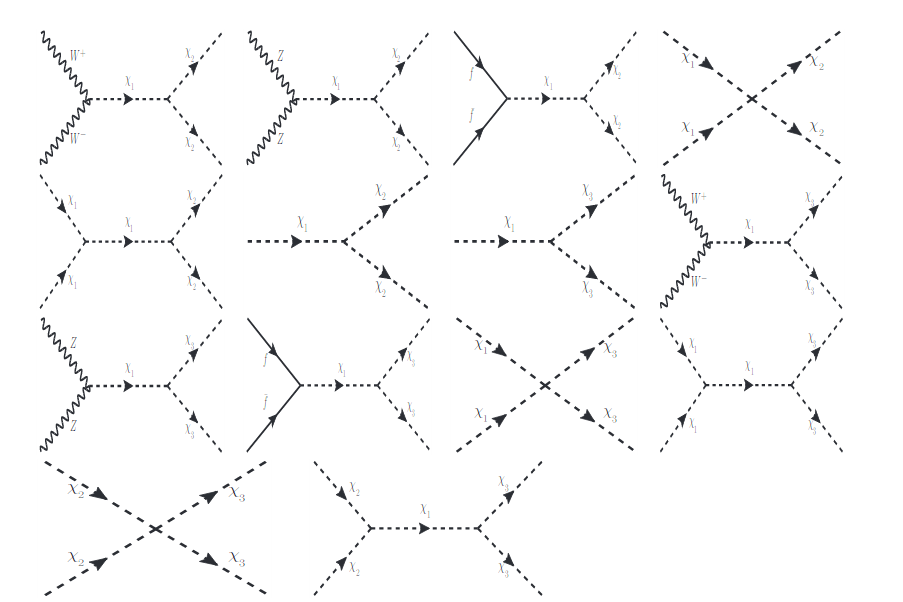}
  \label{fig:3}

\caption{Allowed Feynman diagrams for FIMP-FIMP model}
c. Nonthermal two component dark matter model- Biswas et.al.(2015)
\end{figure}
\section{Results and Discussion}
The plots of the relic density of the dark matter particles for different values of the coupling parameters are shown as follows:\\
\begin{figure}[H]
\centering
  \centering
  \includegraphics[width=9cm]{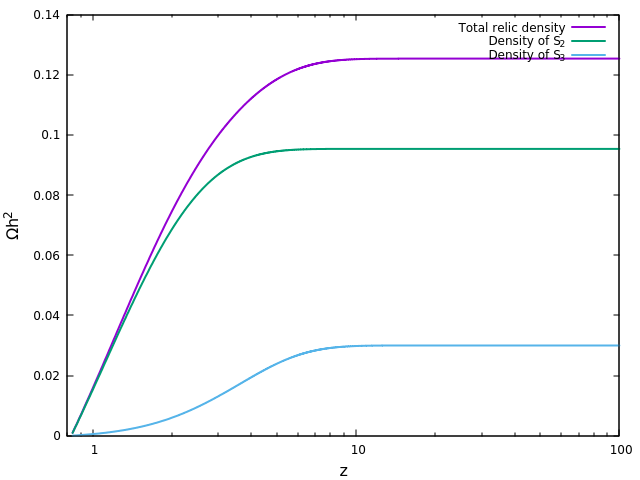}
  \label{fig:5.2}

\caption{Plot showing the fractional contribution of $ S_{2} $ and $ S_{3} $ to the total dark matter relic abundance for the coupling parameters: $ \lambda_{12} = 1.35 \times 10^{-11}, \lambda_{13} = 1.25\times 10^{-9} $}

\end{figure}

\begin{figure}[H]
\centering
\begin{subfigure}{.5\textwidth}
  \centering
  \includegraphics[width=8cm]{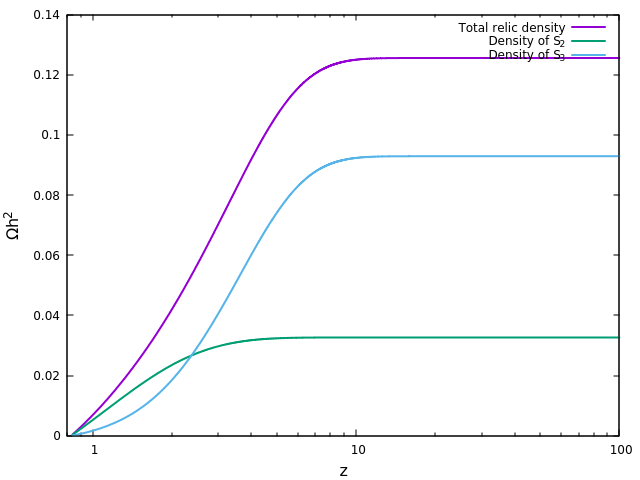}
  \caption{}
  \label{fig:1}
\end{subfigure}%
\begin{subfigure}{.5\textwidth}
  \centering
  \includegraphics[width=8cm]{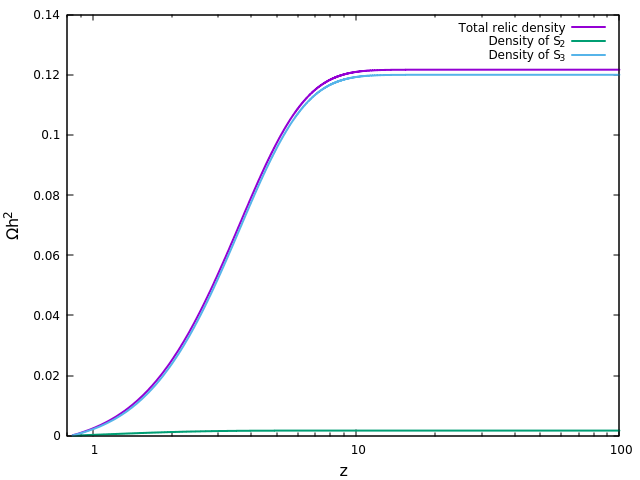}
  \caption{}
  \label{fig:2}
\end{subfigure}
\caption{(a)$ \lambda_{12} = 0.79 \times 10^{-11}, \lambda_{13} = 2.20\times 10^{-9} $\hspace{10pt}  (b)$ \lambda_{12} = 1.8 \times 10^{-12}, \lambda_{13} = 2.5\times 10^{-9} $ Fractional contribution of $ S_{2} $ and $ S_{3} $ to the total dark matter relic abundance for different values of $ \lambda_{12} $ and $ \lambda_{13} $  }
\end{figure}
The parameter space of the coupling parameters are shown below \cite{51}:\\
\begin{figure}[H]
\centering
  \centering
  \includegraphics[width=16cm]{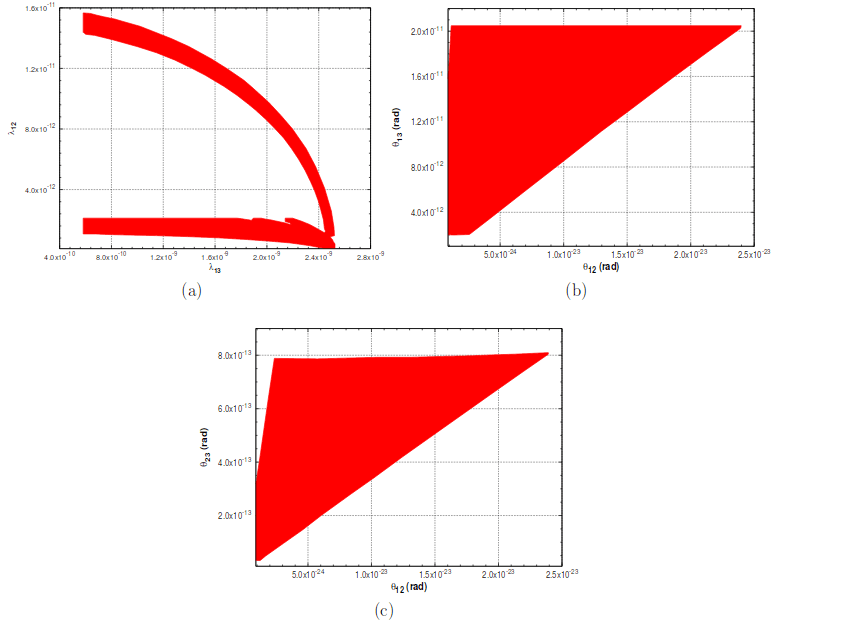}
  \label{fig:3}

\caption{Allowed ranges of model parameters $ \lambda_{12} $, $ \lambda_{13} $, $\theta_{12}$, $\theta_{13}$, $\theta_{23}$  }
\end{figure}

The allowed values of the parameters $\lambda_{12}  $, $\lambda_{13} $ and mixing angles $ \theta_{12},\theta_{13}, \theta_{23} $ obtained from Figure 5.4 (a)-(c) are given below \cite{51}:\\

\begin{center}
\begin{tabu} to 0.9\textwidth { | X[c] | X[c] | X[c] | X[c] | X[c] | }

 \hline
 $ \lambda_{12}$ & $ \lambda_{13}$ & $ \theta_{12}$ & $\theta_{13}$ & $ \theta_{23} $ \\
 \hline
 (0.18-1.6)$ \times $10$ ^{-11} $  & (0.56-2.6)$ \times $10$ ^{-9} $  & (0.1-2.4)$ \times $10$ ^{-23} $ & (0.2-2.0)$ \times $10$ ^{-11} $ & (0.5-8.0)$ \times $10$ ^{-13} $\\
\hline
\end{tabu}
\end{center}
We can see that for different combinations of the model parameters, the fractional relic density of the two FIMP components changes, as shown in Figures 5.2 and 5.3 (a) and (b).
It is ensured that for all the three combinations, the relic density is within the prescribed limit for dark matter set by PLANCK results, which is 0.1172$ \leq \Omega_{DM}h^{2} \leq$0.1226.\\
In the context of dark matter self-interactions, this freedom of being able to tune the parameters and thereby change the fractional contribution of $ S_{2} $ and $ S_{3} $ is particularly important, since dark matter particles of all masses are not capable of undergoing self-intercations, as proved by Campbell et.al. \cite{7}. Only those particles with masses within 0.15 GeV can self-interact, for allowed values of self-interaction cross-sections. Therefore, this model gives me the freedom to set what amount of dark matter in the universe would undergo self-interactions. Simulation results that predict that amount or percentage can hence be used to constrain these parameters of the model.\\
In the next chapters, I describe results of cosmological simulations of self-interacting dark matter and discuss its impact on large-scale cosmology.

\chapter{Cosmological Simulations }

In order to test the impact of dark matter self interactions on large scale structures, I ran cosmological N-body simulations using the parallel TreePM code GADGET-2 developed by Volker Springel for $ \Lambda $CDM simulations and a modified version of GADGET for self-interacting dark matter called the SIDM-GADGET developed by Dr. Jun Koda. \cite{50}. 

It is to be noted here that running large scale cosmological simulations is computationally expensive and requires advanced CPU capabilites, possibly threaded together with 10-12 processors and run parallely. Due to the unavailability of such resources, I had to limit my simulations to a small box of side length 10 Mpc and (32)$ ^{3} $ particles, which is very less for a cosmological simulation. Although, at such a small length scale the fundamental modes are strictly in the non-linear regime, the purpose of my investigation here is to use the simulations to compare $ \Lambda $CDM universe with different self-interacting dark matter models and infer any large-scale changes from them. Therefore, these simulations are run under the sole motivation of comparison between different models alone and are not meant to reveal any precise details of the real universe or match with actual observations.  

\section{Description of N-body codes}
\subsection{GADGET-2}
GADGET-2 is a freely available N-body code for cosmological and hydrodynamic simulations, developed by Volker Springel. \cite{52} The first version of GADGET was released in March 2000 and has since then been widely used for different kinds of simulations like: cosmological structure formation, spherical collapse of self-gravitating spheres of gas and galaxy collisions.\\
It is a TreePM code, which means it distributes the particles in the system in a tree-like structure and uses a Particle Mesh (PM) algorithm to solve long range gravitational forces. The long range forces are approximated by regarding distant groups or clusters of particles as a single unit, whose mass is equal to its centre-of-mass. This algorithm is faster than other algorithms for N-body problems and is able to successfully avoid the pitfalls of other methods with regard to the resolution of short range forces.

\subsection{SIDM-GADGET}
The SIDM-GADGET code for self-interacting dark matter incorporates self-scattering between dark matter particles with a cross-section that is supplied by the user. Self-interactions are treated using a Monte-Carlo approach in which two particles scatter if a random number between (0,1) is less than the local scattering probability within a given simulation time-step given by $ \Delta $t. The scattering probability is given by:
\begin{center}
$ P_{ij} = \rho_{ij}\mid v_{i}-v_{j}\mid \left(\sigma/m\right) \Delta t  $
\end{center} 
where, $ \rho_{ij} $ is the local density of the region, v$ _{i} $ and v$ _{j} $ are the velocities of the two particles considered and $ m $ is the mass of the dark matter particles.\\
The discretised distribution function $ f $ of the particles is given by:\\
\begin{center}
$ f(x,v) = \mathlarger{\mathlarger{\sum}}_{j}mW\left(x-x_{j};r_{j}^{kth}\right) \delta^{3}\left( v-v_{j}\right)  $
\end{center}
where $ W(x,r^{k}) $ is a spline kernel function of size $ r^{k} $.\\
Each macroparticle i, centered at r$ _{i} $, is coarse-grained over a finite spatial patch using the cubic spline kernel $ W(x,r^{k}) $. The local density is defined as:\\
\begin{center}
\begin{large}
 $\rho_{ij}= \int d^{3}r W(x-x_{j},r^{k})W(x-x_{j},r^{k})$ 
\end{large}
\end{center}
Using the SIDM-GADGET code, we can model different dark matter physics, including scenarios where the dark matter particles have a velocity-dependent self-interaction cross-section which is Maxwellian or a power-law velocity distribution. Other models include dark matter interaction via a Yukawa potential, which has been studied in some scenarios. \cite{53}\\
\section{Simulation results and Discussion}
I have run cosmological DM-only simulations for the following dark matter models:
\begin{itemize}
\item $ \Lambda $CDM
\item $ \sigma/m $= 0.47, 1.0, 1.5, 10.0, 300.0 cm$ ^{2} $/g, where $\sigma/m  $ is the DM self scattering cross-section per unit mass.
\item Velocity dependent cross-section - Maxwellian distribution.
\item DM-DM interaction with a Yukawa potential.

\end{itemize}
The halo catalog and halo-mass function described later has been plotted using only the $ \Lambda $CDM and velocity-independent cross-sections. The initial conditions were generated using the publicly available code, N-GenIC by using the Eisenstein and Hu matter power spectrum \cite{54}. \\
The cosmological parameters of the simulations are as follows:
\vspace{1cm}
\begin{center}
\begin{tabu} to 0.8\textwidth { | X[c] | X[c] | X[c] | }
 \hline
 Total matter density ($ \Omega_{0} $) &Total vacuum energy density ($ \Omega_{\Lambda} $) & Hubble parameter \\
 \hline
 0.3  & 0.7  & 0.7 \\ 
\hline
\end{tabu}
\end{center}
\vspace{1cm}

The characteristics of the simuation run are: \\

\begin{center}
\begin{tabu} to 1.0\textwidth { | X[c] | X[c] | X[c] | X[c] | X[c] | }
 \hline
 Starting redshift & Ending redshift & Box size & Number of particles & Mass of each macroparticle \\
 \hline
 10.0  & 0.0  & 10 Mpc & (32)$^{3}$ & 10$^{10}$M$_{\odot} $\\
\hline
\end{tabu}
\end{center}

The following are the snapshots of the runs at z= 10.0, 4.0, 1.0 and 0.0 :\\
\begin{figure}[H]
\centering
  \centering
  \includegraphics[width=18cm,height=4cm]{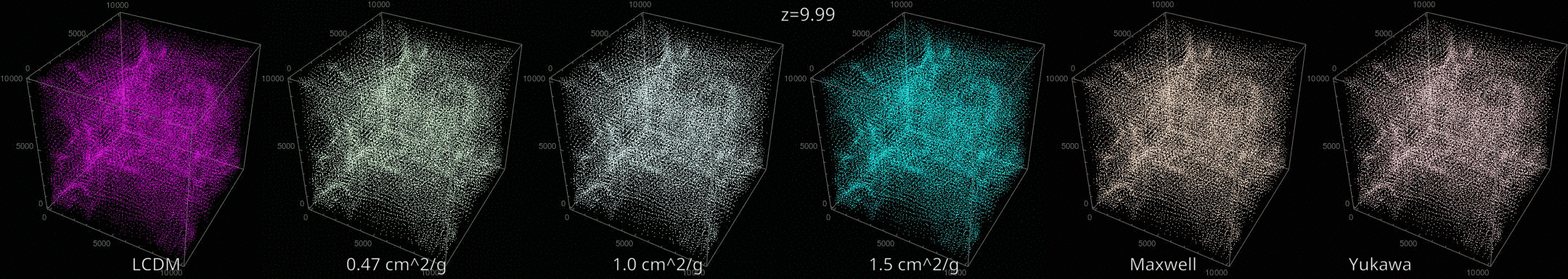}
  \label{fig:3}

\caption{Structure formation at z=9.99 for $\Lambda$CDM and SIDM models}
\end{figure}
\begin{figure}[H]
\centering
  \centering
  \includegraphics[width=18cm,height=4cm]{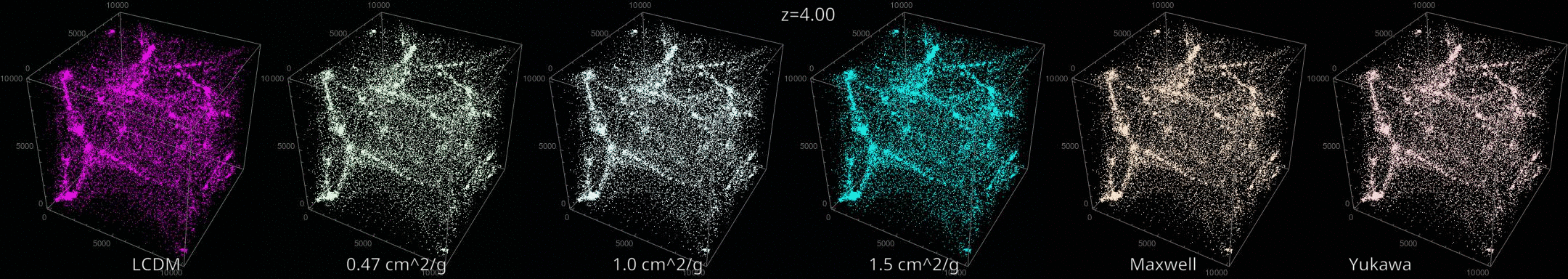}
  \label{fig:3}

\caption{Structure formation at z=4.0 for $\Lambda$CDM and SIDM models}
\end{figure}

\begin{figure}[H]
\centering
  \centering
  \includegraphics[width=18cm,height=4cm]{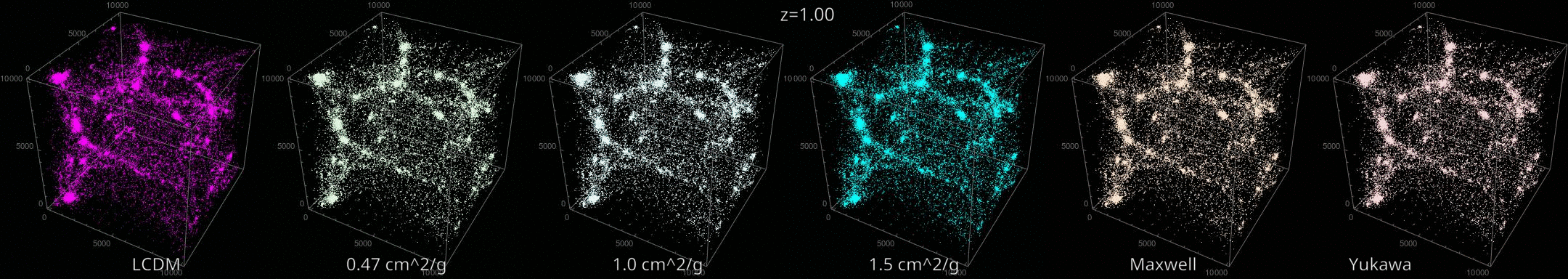}
  \label{fig:3}

\caption{Structure formation at z=1.0 for $\Lambda$CDM and SIDM models}
\end{figure}

\begin{figure}[H]
\centering
  \centering
  \includegraphics[width=18cm,height=4cm]{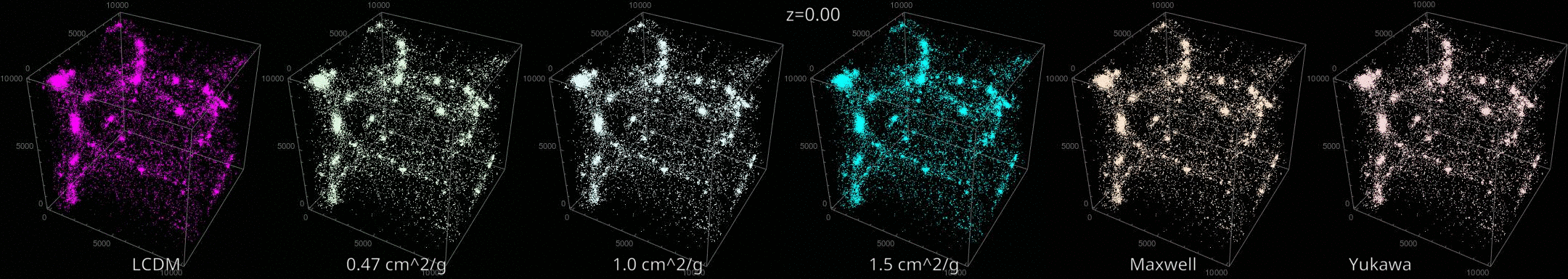}
  \label{fig:3}

\caption{Structure formation at z=0.0 for $\Lambda$CDM and SIDM models}
\end{figure}

As we can see, there is no apparent difference between $ \Lambda $CDM and self interaction models, thereby confirming the fact that DM self interactions do not affect structures at large scales and thereby retains the success of $ \Lambda $CDM at large scales.
However, a separate simulation at only z=0 for each of these models revealed small fluctuations in the density and size of the halos. Figures 6.5-6.7 show the snapshots of the individual models zoomed at a particular region of the box. However to be able to see the change better, click the link and watch the simulation \href{https://drive.google.com/open?id=1yO6Rjf8fEScMnpuSPB31AOKWydROSwmq}{movie}\\
 
\begin{figure}[H]
\centering
\begin{subfigure}{.5\textwidth}
  \centering
  \includegraphics[width=6.0cm]{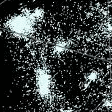}
  \caption{}
  \label{fig:6}
\end{subfigure}%
\begin{subfigure}{.5\textwidth}
  \centering
  \includegraphics[width=5.9cm]{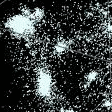}
  \caption{}
  \label{fig:2}
\end{subfigure}
\caption{$ \Lambda $CDM \hspace{150pt} 0.47 cm$ ^{2} $/g}
\end{figure}

\begin{figure}[H]
\centering
\begin{subfigure}{.5\textwidth}
  \centering
  \includegraphics[width=6.0cm]{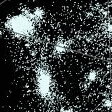}
  \caption{}
  \label{fig:1}
\end{subfigure}%
\begin{subfigure}{.5\textwidth}
  \centering
  \includegraphics[width=5.9cm]{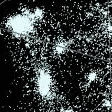}
  \caption{}
  \label{fig:2}
\end{subfigure}
\caption{1.0 cm$ ^{2} $/g \hspace{150pt} 1.5 cm$ ^{2} $/g  }
\end{figure}

\begin{figure}[H]
\centering
\begin{subfigure}{.5\textwidth}
  \centering
  \includegraphics[width=6.0cm]{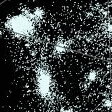}
  \caption{}
  \label{fig:1}
\end{subfigure}%
\begin{subfigure}{.5\textwidth}
  \centering
  \includegraphics[width=5.9cm]{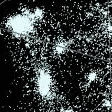}
  \caption{}
  \label{fig:2}
\end{subfigure}
\caption{Maxwell distribution \hspace{100pt} Yukawa interaction  }
\end{figure}
 
Now that it is established that dark matter self interactions do not change the number of halos at cosmological scales, with possible changes at the sub-halo scales, I have plotted the halo mass function and the halo catalog using a Friends-of-Friends (FOF) algorithm as described in the next section.
 
\section{Halo mass function}
The halo mass function is a plot of the cumulative number of halos vs. mass obtained from a cosmological simulation. Several algorithms for finding halos exist, a popular method being the Friends-of-Friends which has been implemented here. I have used a halofinder code to identify the halos and to obtain the cumulative halo mass function. The description of the Friends-of-Friends algorithm is given below:

\subsection{Friends-of-Friends(FOF) algorithm}
The FOF algorithm is used to identify structures in numerical simulations based on the physical proximity of the particles. The single free parameter in the FOF approach is the linking length which is often quoted in terms of the mean inter-particle separation. Two particles are called `friends' if the distance between them is less than the linking length and if this two particles have more friends associated with them, they form a cluster. A cluster consisting of some specified number of particles form a halo. In this work, I have set the condition that a cluster consisting of 30 particles form a halo and the linking length is set as 0.2 Kpc. Using this, a linked list consisting of the masses of the identified halos and the number of particles constituting them are generated, which is used to generate the halo catalog. 
 
\subsection{Results}
\begin{figure}[H]
\centering
  \centering
  \includegraphics[width=12cm]{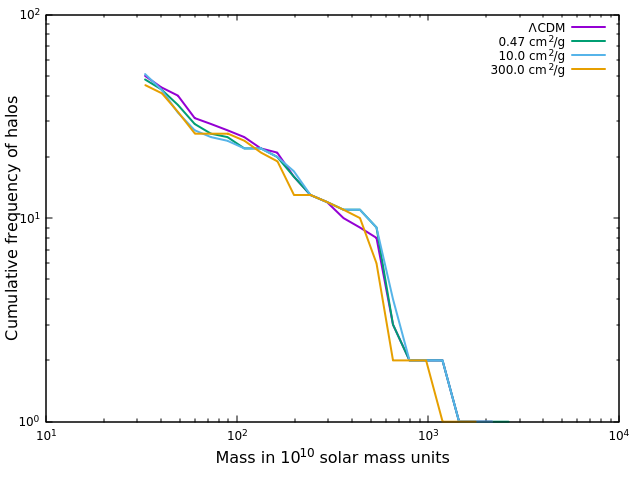}
  \label{fig:3}

\caption{Cumulative number of halos vs. mass}
\end{figure}
We can see from the figure above that there is very little change in the total number of halos for even very extreme and unrealistic cross-sections like 10 and 300 cm$ ^{2} $/g, implying that self interacting dark matter preserves the large-scale success of $ \Lambda $CDM with regard to halo population. The halo catalog showing the number of halos corresponding to each mass bin is shown below:

\begin{figure}[H]
\centering
  \centering
  \includegraphics[width=12cm]{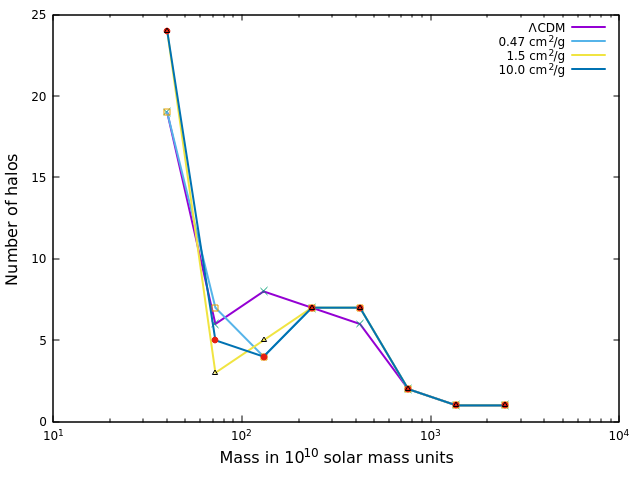}
  \label{fig:3}

\caption{Number of halos vs. mass}
\end{figure}

We can see that the number of halos corresponding to each mass bin follows a more or less similar pattern for each of the cross-sections, despite small fluctuations. A similar analysis with better mass resolution and more number of particles would be interesting since that would help indicate whether there is some mass scale where the effect of self-interactions on number of halos is more pronounced. The plot above suggests that the fluctuation is more on small mass scales (60-110) $ \times $10$ ^{10}  $ M$ _{\odot} $. A larger simulation would reveal whether or not this is due to some statistical effect or an underlying physics.

\section{Subhalo population}
I made an attempt to find out the subhalo population from the simulation snapshots using the code Amiga Halo Finder (AHF) which is freely available online. Having run the code, I found no subhalos. This is expected since (32)$ ^{3} $ particles and 10 Mpc box size is too small for a cosmological simulation. I plan to run a larger simulation with (512)$ ^{3} $ particles and use AHF to calculate the subhalo population and match my results with the predictions of self-interacting dark matter.

\chapter{ETHOS- An Effective Theory of Structure Formation}

In this chapter, I shall briefly describe the newly developed ETHOS framework by Mark Vogelsberger and his group \cite{55} and the impact on the matter power spectrum of a particular ETHOS toy-model.

\section{What is ETHOS?}
Until very recently, dark matter physics research had been undertaken separately by two communities: particle physicists, who primarily work on the phenomenological aspects of dark matter physics, like the designing of dark matter models and working out the nature of their interactions with observed matter; and cosmologists who use the prevalent knowledge of dark matter physics to address several important cosmological questions like large-scale structure formation and evolution, galaxy evolution, mergers etc.

ETHOS represents a shift in the paradigm in the sense that it directly tries to investigate how dark matter particle physics models affect the growth of structure by putting together both particle physics models and cosmological parameters into a single framework. It is a mapping of particle physics theories into physical effective ETHOS parameters that shape the linear matter power spectrum and transfer function. As desribed in the first ETHOS paper, \cite{55} ``The usefulness of such a framework is clear: all dark matter particle models that map to a given effective theory can be simulatneously constrained by comparing a single ETHOS simulation to observations at no extra cost or effort." In order to test the ETHOS paradigm, the publicly available code ETHOS-CAMB \cite{55} has been used for the work described below:
%SC cite the paper
\section{An Example}
Following is an example of a toy model, whose impact on the linear matter power spectrum  is studied using the ETHOS framework. The details of this work is described in the paper. \cite{56}. Here I have reproduced a very small part of that work.

The DM particle in this model is a Dirac fermion $ \chi $ that interacts with a nearly massless sterile neutrino, which plays the role of dark radiation through a massive vector boson $ \phi_{\mu} $.\\
The interaction Lagrangian is given by:\\
\begin{center}
$ \mathcal{L}_{int} = -\dfrac{1}{2}m_{\phi}^{2}\phi_{\mu}\phi^{\mu} - \dfrac{1}{2}m_{\chi}\bar{\chi}\chi - g_{\chi}\phi_{\mu}\bar{\chi}\gamma^{\mu}\chi - \dfrac{1}{2}g_{\nu}\phi_{\mu}\bar{\nu_{s}}\gamma^{\mu}\nu_{s} $\\
\end{center}
The non-zero ETHOS parameters in this model are as follows:\\

$ \omega_{DR} = 1.35 \times 10^{-6}\left( \dfrac{\xi}{0.5}\right)^{4} $\\
\vspace{10pt}
$ a_{4} = \left(1+z_{D}\right)^{4}\dfrac{3\pi}{2}\dfrac{g_{\chi}^{2}g_{\nu}^{2}}{m_{\phi}^{4}}\dfrac{\rho_{c}/h^{2}}{m_{\chi}}\Big(\dfrac{310}{441}\Big)\xi^{2}T^{2}_{CMB,0}  $\\
\vspace{10pt}
$ = 0.6 \times 10^{5}\Big(\dfrac{g_{\chi}}{1}\Big)^{2}\Big(\dfrac{g_{\nu}}{1}\Big)^{2}\Big(\dfrac{0.468 MeV}{m_{\phi}}\Big)^{4}\Big(\dfrac{2 TeV}{m_{\chi}}\Big)\Big(\dfrac{\xi}{0.5}\Big)^{2} Mpc^{-1}  $\\
\vspace{10pt}
$ \alpha_{l \geq 2} = \dfrac{3}{2} $\\

The mapping from particle physics parameters to effective ETHOS parameters is therefore given by:\\
\begin{center}
$\lbrace m_{\chi}, m_{\phi}, g_{\chi}, g_{\nu}, \xi, \eta_{\chi}, \eta_{\nu_{s}} \rbrace \longrightarrow \lbrace \omega_{DR}, a_{4}, \alpha_{l \geq 2 = \frac{3}{2}} \rbrace$
\end{center}
Here, $ \omega_{DR} $ is the relic density ($ \Omega_{DR}h^{2} $) of the dark radiation component, $ a_{4} $ is an ETHOS parameter related to the fourth power of $ z_{D} $ which is the redshift at which the dark matter opacity becomes equal to the conformal Hubble rate, $ \alpha_{l} $ are l-dependent parameters related to the DM-DR scattering cross-section. $ \eta_{\chi} $ and $ \eta_{\nu_{s}} $ are dark matter and sterile neutrino spin and color degeneracy factors respectively. $ \xi $ is the ratio of dark matter to CMB temperature at redshift 0.
While deriving the perturbation equations for a dark matter-dark radiation model it is assumed that the particle physics parameters (coupling and masses) enter the Boltzmann equation through opacity co-efficients ($ \dot{\kappa} $) which are related to the mean free path of the particle by:
\begin{center}
$ -\dot{\kappa} \simeq 1/\lambda \simeq (n \sigma) $
\end{center}
where $ n $ is the number density of the targets and $ \sigma $ is the scattering cross-section.

In ETHOS parameterization, the dark matter opacity parameter, $ \dot{\kappa_{\chi}} $ scales with redshift, according to the relation:
\begin{center}
$ \dot{\kappa_{\chi}} = -\dfrac{4}{3}\Omega_{DR}h^{2}\mathlarger{\mathlarger{\sum}}_{n}a_{n}\dfrac{(1+z)^{n+1}}{(1+z_{D})^{n}} $
\end{center} 
Putting the ETHOS parameters as input in the ETHOS-CAMB code, the following matter power spectrum was obtained for z=124:\\
\begin{center}
\begin{figure}[H]
\centering
  \centering
  \includegraphics[width=15cm]{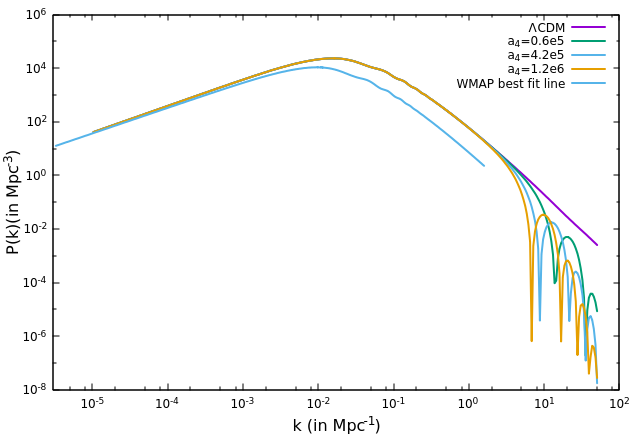}
  \label{fig:3}

\caption{Matter power spectrum for different values of parameter $ a_{4} $ compared with $ \Lambda $CDM at z=0. Also, the best fit matter power spectrum curve from WMAP data has been plotted for comparison. \cite{74} }
\end{figure}
\end{center}
The oscillatory and damping features relative to $ \Lambda $CDM as evident in the power spectrum are signatures of Dark Acoustic Oscillations (DAO) \cite{57} and Silk damping.
Dark Acoustic Oscillations arise in the early universe when DM and DR were tightly coupled and the collapse of DM density perturbations are prevented by the pressure created by the DR fluid.\\
%SC Kaustav's comment on the amplitude and constraints from current observations.  Add that part here. 
We can see from Figure 7.1 that the oscillatory DAO features arise at length scales smaller than 1 Mpc. The observed matter power at such small scales are heavily suppressed, as can be seen from the WMAP data. \cite{62} Therefore, the model cannot be constrained from the WMAP best fit power spectrum. Lyman alpha or weak lensing data at smaller length scales may be used for that purpose. \\

% body of thesis comes here

\chapter{Conclusion}

\label{ch:conclusions}

The aim of this project was to establish a theoretical particle physics model for dark matter that can interact with itself and to study its effect on the large scale structure of the universe. After having constructed the models, the relic density for the FIMP-FIMP model was calculated. I have then tried to investigate the halo population of different dark matter models and found no observable large scale difference. However minor changes in the subhalo population exists, as apparent from the simulation snapshots. While the small scale simulations do not reflect any changes, in order to quantify the change, much larger simulations are needed on box size of large length scale, so that the analysis can be done in the linear regime. \\
The connection between the theoretical model and the cosmological simulations was done by tweaking certain interaction parameters in the potential of the FIMP-FIMP dark matter system, so that the maximum contribution to the total dark matter content of the universe came from the lighter FIMP particle which could undergo self-interactions, following the prescription set by Campbell et.al. (2015). This work can therefore be thought of as an attempt to directly use a particle physics model to understand a cosmological problem, which is also the philosophy behind the newly developed elaborate ETHOS framework. I have in conclusion reproduced a small portion of the results of the paper \cite{56} in order to test the ETHOS scheme on my own and explore the vast posibilities of this new scheme.    

% Rather than itemize write out the future plans and say how you see them to add to your understanding of the model.  
\section{Future Work}
In the future, I plan to large larger simulations, possibly with (512)$ ^{3} $ particles and box size of 150 Mpc or more and check for changes in subhalo population for different dark matter self interaction models using AHF. Also, the change in density profiles of subhalos and rotation curves of galaxies can be inspected for and compared with $ \Lambda $CDM. Further, simulations of isolated halos can give us an idea about the change in ellipticity of halos predicted for many self-interacting DM models, including Yoshida et.al.(2000) \cite{45}. Besides, the same checks can be made for velocity dependent self-interaction cross-sections and compared with $ \Lambda $CDM. It may hence be possible to derive some length scale for which self-interaction cross-sections show a velocity dependence and otherwise.\\
Further work can also be done with ETHOS and the full implications of different ETHOS models on cosmological structure formation can be studied and the models can be simulatenously constrained from observations.\\  
\addcontentsline{toc}{chapter}{Bibliography}
\bibliographystyle{alpha}
\bibliography{bibliography/bibliography}

\begin{thebibliography}{}
\bibitem{7} Robyn Campbell, Stephen Godfrey, Heather E. Logan, Andrea D. Peterson,and Alexandre Poulin \href{https://journals.aps.org/prd/abstract/10.1103/PhysRevD.92.055031}{https://journals.aps.org/prd/abstract/10.1103/PhysRevD.92.055031}
\bibitem{55} Francis-Yan  Cyr-Racine, Kris  Sigurdson, Jesus Zavala Torsten  Bringmann, Mark  Vogelsberger, and  Christoph  Pfrommer \href{https://arxiv.org/abs/1512.05344}{https://arxiv.org/abs/1512.05344}
\bibitem{1} Planck Collaboration \href{https://doi.org/10.1051/0004-6361/201321591}{https://doi.org/10.1051/0004-6361/201321591}
\bibitem{2} Planck Collaboration \href{https://doi.org/10.1051/0004-6361/201525830}{https://doi.org/10.1051/0004-6361/201525830}
\bibitem{3} Jounghun Lee, Eiichiro Komatsu \href{http://iopscience.iop.org/article/10.1088/0004-637X/718/1/60/meta}{http://iopscience.iop.org/article/10.1088/0004-637X/718/1/60/meta}
\bibitem{4} Rubin V.C., Ford Jr. W.K., Thonnard N., 1980  \href{http://adsabs.harvard.edu/abs/1980ApJ...238..471R}{http://adsabs.harvard.edu/abs/1980ApJ...238..471R}
\bibitem{5} Håkon Dahle \href{http://iopscience.iop.org/article/10.1086/508654/meta}{http://iopscience.iop.org/article/10.1086/508654/meta}
\bibitem{6} Fritz Zwicky \href{http://adsabs.harvard.edu/doi/10.1086/143864}{http://adsabs.harvard.edu/doi/10.1086/143864}

\bibitem{8} Horace W. Babcock \href{http://adsabs.harvard.edu/abs/1939LicOB..19...41B}{http://adsabs.harvard.edu/abs/1939LicOB..19...41B}
\bibitem{9} Rubin V.C., Ford Jr.W.K., Thonnard N., 1980, \href{http://adsabs.harvard.edu/abs/1980ApJ...238..471R}{http://adsabs.harvard.edu/abs/1980ApJ...238..471R}
\bibitem{10} Loewenstein, M. \href{http://adsabs.harvard.edu/full/1991BAAS...23..891L}{http://adsabs.harvard.edu/full/1991BAAS...23..891L}
\bibitem{11} Daylan, Finkbeiner \href{https://doi.org/10.1016/j.dark.2015.12.005}{https://doi.org/10.1016/j.dark.2015.12.005}
\bibitem{12} J.Petrovic, P.D. Serpico and G. Zaharijas, JCAP 1502, no. 02, 023 (2015) \href{https://arxiv.org/pdf/1501.02666.pdf}{https://arxiv.org/pdf/1501.02666.pdf}
\bibitem{13} M.S. Boucenna and S.Profumo \href{https://journals.aps.org/prd/abstract/10.1103/PhysRevD.84.055011}{https://journals.aps.org/prd/abstract/10.1103/PhysRevD.84.055011}
\bibitem{14} A. Alves, S. Profumo, F.S. Queiroz and W. Shepherd, Phys.Rev.D 90, no.11, 115003 (2014) \href{https://arxiv.org/pdf/1403.5027.pdf}{https://arxiv.org/pdf/1403.5027.pdf}
\bibitem{15} A. Berlin, D. Hooper and S.D. McDermott, Phys.Rev.D 89, 115022 (2014) \href{https://journals.aps.org/prd/abstract/10.1103/PhysRevD.89.115022}{https://journals.aps.org/prd/abstract/10.1103/PhysRevD.89.115022}
\bibitem{16} Markevitch. et.al.(2002) \href{http://iopscience.iop.org/article/10.1086/339619/fulltext/15847.text.html}{http://iopscience.iop.org/article/10.1086/339619/fulltext/15847.text.html}
\bibitem{17} R. Krall, M. Reece and T. Roxlo, JCAP 1409, 007 (2014) \href{http://iopscience.iop.org/article/10.1088/1475-7516/2014/09/007/meta}{http://iopscience.iop.org/article/10.1088/1475-7516/2014/09/007/meta}
\bibitem{18} J.C. Park, S.C. Park and K. Kong, Phys.Lett.B 733, 217 (2014) \href{https://doi.org/10.1016/j.physletb.2014.04.037}{https://doi.org/10.1016/j.physletb.2014.04.037}
\bibitem{19} M.T. Frandsen, F. Sannino, I. M. Shoemaker and O. Svendsen, JCAP 1405, 033 (2014) \href{https://doi.org/10.1016/j.physletb.2014.04.037}{https://doi.org/10.1016/j.physletb.2014.04.037} 
\bibitem{20} Teresa Marrodan Undagoitia and Ludwig Rauch \href{http://iopscience.iop.org/article/10.1088/0954-3899/43/1/013001/meta}{http://iopscience.iop.org/article/10.1088/0954-3899/43/1/013001/meta}
\bibitem{21} Jianglai Liu, Xun Chen, and Xiangdong Ji \href{https://arxiv.org/pdf/1709.00688.pdf}{https://arxiv.org/pdf/1709.00688.pdf}
\bibitem{22} Chung-Lin Shan \href{https://www.nature.com/articles/nphys4039}{doi:10.1038/nphys4039}
\bibitem{23} LUX Collaboration \href{https://journals.aps.org/prl/abstract/10.1103/PhysRevLett.112.091303}{PhysRevLett.112.091303}
\bibitem{24} LUX-ZEPLIN Collaboration \href{https://arxiv.org/pdf/1802.06039.pdf}{https://arxiv.org/pdf/1802.06039.pdf}
\bibitem{25} XENON-100 Collaboration \href{https://journals.aps.org/prd/abstract/10.1103/PhysRevD.94.122001}{Phys. Rev. D 94, 122001}
\bibitem{26} DarkSide Collaboration \href{https://arxiv.org/pdf/1802.07198.pdf}{https://arxiv.org/pdf/1802.07198.pdf}
\bibitem{27} SuperCDMS Collaboration \href{https://journals.aps.org/prl/abstract/10.1103/PhysRevLett.120.061802}{Phys. Rev. Lett. 120, 061802 }
\bibitem{28} CDEX \href{http://iopscience.iop.org/article/10.1088/1674-1137/37/12/126002/meta}{http://iopscience.iop.org/article/10.1088/1674-1137/37/12/126002/meta}
\bibitem{29} CoGENT Collaboration \href{https://journals.aps.org/prd/abstract/10.1103/PhysRevD.88.012002}{Phys. Rev. D 88, 012002}
\bibitem{30} CRESST Collaboration \href{https://arxiv.org/pdf/1711.07692.pdf}{https://arxiv.org/pdf/1711.07692.pdf}
\bibitem{31} PICO Collaboration \href{https://journals.aps.org/prl/abstract/10.1103/PhysRevLett.118.251301}{Phys. Rev. Lett. 118, 251301 }
\bibitem{32} XENON 1T Collaboration \href{https://journals.aps.org/prl/abstract/10.1103/PhysRevLett.119.181301}{Phys. Rev. Lett. 119, 181301}
\bibitem{69} Teresa Marrodán Undagoitia and Ludwig Rauch  \href{http://iopscience.iop.org/article/10.1088/0954-3899/43/1/013001/meta}{J. Phys. G43 (2016) no.1, 013001}
\bibitem{70} Bowman, J. D., Rogers, A.E.E., Monsalve, R. A., Mozdzen, T.J. and Mahesh, N. Nature 555, 67–70 (2018) \href{https://www.nature.com/articles/nature25792}{doi:10.1038/nature25792}
\bibitem{71} Julio Amare, Susana Cebrian et.al. \href{https://arxiv.org/pdf/1601.01184.pdf}{https://arxiv.org/pdf/1601.01184.pdf}
\bibitem{72} K Fushimi, S Nakayama, R Orito, R Sugawara, Y Awatani, H Ejiri, T Shima, R Hazama, K Inoue, H Ikeda \href{http://iopscience.iop.org/article/10.1088/1742-6596/469/1/012011}{http://iopscience.iop.org/article/10.1088/1742-6596/469/1/012011}
\bibitem{73} \href{https://www.nature.com/articles/d41586-018-03991-y}{https://www.nature.com/articles/d41586-018-03991-y}
\bibitem{34} Howard Baer, Ki-Young Choi, Jihn E. Kim \href{https://doi.org/10.1016/j.physrep.2014.10.002}{https://doi.org/10.1016/j.physrep.2014.10.002}
\bibitem{35} Hiroshi Okada, Yuta Orikasa and Takashi Toma \href{https://journals.aps.org/prd/abstract/10.1103/PhysRevD.93.055007}{Phys. Rev. D 93, 055007}
\bibitem{36} Bernal et.al. \href{https://doi.org/10.1142/S0217751X1730023X}{https://doi.org/10.1142/S0217751X1730023X}
\bibitem{37} Douglas Scott \href{https://arxiv.org/pdf/1804.01318.pdf}{https://arxiv.org/pdf/1804.01318.pdf}
\bibitem{38} Antonino Del Popolo and Morgan Le Delliou \href{https://doi.org/10.3390/galaxies5010017}{https://doi.org/10.3390/galaxies5010017}
\bibitem{39} Navarro, J.F.; Frenk, C.S.; White, S.D.M.  \href{http://xxx.lanl.gov/pdf/astro-ph/9508025v1}{http://xxx.lanl.gov/pdf/astro-ph/9508025v1}
\bibitem{40} Flores, R.A.; Primack, J.R. \href{http://xxx.lanl.gov/pdf/astro-ph/9402004v1}{http://xxx.lanl.gov/pdf/astro-ph/9402004v1}
\bibitem{41} Boylan-Kolchin, M.; Bullock, J.S.; Kaplinghat, M.  \href{http://xxx.lanl.gov/pdf/1103.0007v2}{http://xxx.lanl.gov/pdf/1103.0007v2}
\bibitem{42} Moore, B.; Quinn, T.; Governato, F.; Stadel, J.; Lake, G. \href{http://xxx.lanl.gov/pdf/astro-ph/9903164v1}{http://xxx.lanl.gov/pdf/astro-ph/9903164v1}
\bibitem{43} J.S. Bullock \href{https://arxiv.org/pdf/1009.4505.pdf}{https://arxiv.org/pdf/1009.4505.pdf} 
\bibitem{46} David N. Spergel and Paul J. Steinhardt \href{https://journals.aps.org/prl/abstract/10.1103/PhysRevLett.84.3760}{https://journals.aps.org/prl/abstract/10.1103/PhysRevLett.84.3760}
\bibitem{44} L. Goodenough and D. Hooper, arXiv:0910.2998 [hep-ph] 
\bibitem{45} Naoki Yoshida,Volker Springel, Simon D.M. White \href{http://iopscience.iop.org/article/10.1086/317306/pdf}{http://iopscience.iop.org/article/10.1086/317306/pdf}
\bibitem{47}  F. Kahlhoefer, K. Schmidt-Hoberg, J. Kummer and S. Sarkar, Mon. Not. Roy. Astron. Soc.452, no. 1, L54 (2015).
\bibitem{48} Harvey et.al. \href{http://science.sciencemag.org/content/347/6229/1462}{DOI: 10.1126/science.1261381}
\bibitem{49} M. Kaplinghat, S. Tulin, and H.-B. Yu (2015) \href{https://journals.aps.org/prl/abstract/10.1103/PhysRevLett.116.041302}{Phys. Rev. Lett. 116, 041302}
\bibitem{50} Dr. Jun Koda \href{http://adsabs.harvard.edu/abs/2017ascl.soft03007K}{http://adsabs.harvard.edu/abs/2017ascl.soft03007K}
\bibitem{51} Anirban Biswas, Debasish Majumdar, Probir Roy \href{https://doi.org/10.1007/JHEP04(2015)06}{https://doi.org/10.1007/JHEP04(2015)06}
\bibitem{60} Dutta Banik, A., Pandey, M., Majumdar, D. et al. Eur. Phys. J. C (2017) 77: 657. \href{https://doi.org/10.1140/epjc/s10052-017-5221-y}{https://doi.org/10.1140/epjc/s10052-017-5221-y}
\bibitem{52}  Volker Springel \href{https://doi.org/10.1111/j.1365-2966.2005.09655.x}{https://doi.org/10.1111/j.1365-2966.2005.09655.x}
\bibitem{53} M.H. Chan \href{http://iopscience.iop.org/article/10.1088/2041-8205/769/1/L2/meta}{http://iopscience.iop.org/article/10.1088/2041-8205/769/1/L2/meta}
\bibitem{54} Daniel J. Eisenstein and Wayne Hu \href{http://iopscience.iop.org/article/10.1086/306640/meta}{http://iopscience.iop.org/article/10.1086/306640/meta}
\bibitem{56} Subinoy Das, Rajesh Mondal, Vikram Rentala, Srikanth Suresh \href{https://arxiv.org/pdf/1712.03976.pdf}{https://arxiv.org/pdf/1712.03976.pdf}
\bibitem{74} WMAP cosmological parameters \href{https://lambda.gsfc.nasa.gov/product/map/dr2/params/lcdm_wmap.cfm}{LAMBDA- Data Products}
\bibitem{57} Francis-Yan  Cyr-Racine, Roland de Putter, and Alvise  Raccanelli \href{https://journals.aps.org/prd/abstract/10.1103/PhysRevD.89.063517}{Phys. Rev. D 89, 063517}
\bibitem{58} E. Bulbul, M. Markevitch, A. Foster, R. K. Smith, M. Loewenstein and S. W. Randall, Astrophys. J.789, 13 (2014)
\bibitem{59} Tansu Daylan, Douglas P. Finkbeiner, Dan Hooper, Tim Linden, Stephen K. N. Portillo, Nicholas L. Rodd, and Tracy R. Slatyer \href{https://doi.org/10.1016/j.dark.2015.12.005}{https://doi.org/10.1016/j.dark.2015.12.005}
\bibitem{61} Ran Huo, Manoj Kaplinghat, Zhen Pan, and Hai-Bo Yu \href{https://arxiv.org/pdf/1709.09717.pdf}{https://arxiv.org/pdf/1709.09717.pdf}
\bibitem{62} Andrew Robertson, Richard Massey, Vincent Eke, Richard Bower \href{https://doi.org/10.1093/mnras/stv1805}{https://doi.org/10.1093/mnras/stv1805}
\bibitem{63} Jonathan L. Feng, Manoj Kaplinghat, Huitzu Tu, and Hai-Bo Yu \href{http://iopscience.iop.org/article/10.1088/1475-7516/2009/07/004/meta}{http://iopscience.iop.org/article/10.1088/1475-7516/2009/07/004/meta}
\bibitem{64} Miguel Rocha, Annika H. G. Peter, James S. Bullock, Manoj Kaplinghat, Shea Garrison-Kimmel, Jose Onorbe and Leonidas A. Moustakas \href{https://doi.org/10.1093/mnras/sts514}{https://doi.org/10.1093/mnras/sts514}
\bibitem{65} Scott W. Randall, Maxim Markevitch, Douglas Clowe, Anthony H. Gonzalez, and Marusa Bradac \href{http://iopscience.iop.org/article/10.1086/587859/meta}{http://iopscience.iop.org/article/10.1086/587859/meta}
\bibitem{66} Anirban Biswas, Debasish Majumdar \href{https://doi.org/10.1007/s12043-012-0478-z}{https://doi.org/10.1007/s12043-012-0478-z}
\bibitem{67} Michel H.G. Tytgat \href{https://arxiv.org/pdf/1012.0576.pdf}{https://arxiv.org/pdf/1012.0576.pdf}
\bibitem{68} J. A. Casas, G.A. Gomez Vargas, J. M. Moreno, J. Quilis and R. Ruiz de Austri \href{https://arxiv.org/pdf/1711.10957.pdf}{https://arxiv.org/pdf/1711.10957.pdf}
\end{thebibliography}

\appendix
% appendices come here
%\fi

\end{document}